\documentclass[fleqn,10pt]{wlscirep}
\usepackage[utf8]{inputenc}
\usepackage[T1]{fontenc}

\usepackage{subfiles}
\usepackage{multirow}
\usepackage{tikz} 

\usepackage{amsthm,amsmath,amssymb}
\usepackage{mathptmx}
\usepackage{mathrsfs}
\usepackage{ulem}
\DeclareMathAlphabet{\mathcal}{OMS}{cmsy}{m}{n}  
\usepackage{ulem} 
\usepackage{xpatch}
\usepackage[toc,page]{appendix}
\usepackage{url}
\usepackage{texpower}
\usepackage{amsmath}
\usepackage{booktabs}
\usepackage{makecell}
\usepackage{colortbl}
\usepackage{subcaption}
\usepackage[table]{xcolor}
\definecolor{tableHead}{RGB}{216,214,194}
\definecolor{tableContent}{RGB}{235,234,222}
\usepackage{tabularx}

\usepackage{lineno} 

\definecolor{modification}{RGB}{0,0,0}  

\definecolor{cred}{HTML}{FF6B6B}
\definecolor{cyellow}{HTML}{FEC260}
\definecolor{cgreen}{HTML}{6BCB77}
\definecolor{cgreen}{HTML}{70AD47}
\definecolor{cblue}{HTML}{4D96FF}
\definecolor{cpurple}{HTML}{2A0944}
\definecolor{ggray}{RGB}{127,127,127}
\definecolor{aliceblue}{rgb}{0.94, 0.97, 1.0}

\usepackage{makecell}   

\title{
 Navigating Chemical-Linguistic Sharing Space with Heterogeneous Molecular Encoding
}
\author[1,2,3,4,$\star$]{Liuzhenghao Lv}
\author[1,2,3,$\star$]{Hao Li}
\author[1,3,4]{Yu Wang}
\author[1]{Zhiyuan Yan}
\author[1]{Zijun Chen}
\author[1]{Zongying Lin}
\author[1,2,3,$\dagger$]{Li Yuan}
\author[1,2,3,$\dagger$]{Yonghong Tian}
\affil[1]{\textit{School of Electronic and Computer Engineering, Peking University Shenzhen Graduate School, Shenzhen, 518055, China}}
\affil[2]{\textit{Peng Cheng Laboratory, Shenzhen, 518000, China}}
\affil[3]{\textit{AI for Science (AI4S)-Preferred Program, Peking University Shenzhen Graduate School, Shenzhen, 518055, China}}
\affil[4]{\textit{School of Computer Science, Peking University, Beijing, 10091, China}}
\affil[$\star$]{\textit{These authors contributed equally to this work}}
\affil[$\dagger$]{Corresponding authors: yuanli-ece@pku.edu.cn, yhtian@pku.edu.cn}

\begin{abstract}

Chemical language models~(CLMs) are prominent for their effectiveness in exploring chemical space and enabling molecular design and engineering. 
However, while exploring chemical-linguistic space, CLMs suffer from the semantic gap between natural language and molecular representations. This challenge is primarily due to the inherent modeling differences between molecules and texts: molecules operate unified modeling to learn chemical space, while natural language sequentially models their semantic space. Additionally, the limited availability of high-quality text-to-molecule datasets further exacerbates this challenge.
To address the problem, we first verified the information bias in molecular representations from different perspectives. 
We then developed the Heterogeneous Molecular Encoding~(HME) framework, a unified molecular encoder compressing the molecular features from fragment sequence, topology, and conformation with Q-learning.
To better model chemical-linguistic space, we further constructed the MCMoD dataset, which contains over one million molecules with various conditions, including properties, fragments, and descriptions. 
Experimentally, HME promotes CLMs to achieve chemical-linguistic sharing space exploration from two aspects: (1) chemical space exploration with linguistic guidance, where HME achieves significant improvements~(+8.9\% FCD) for molecular design in multiple constraints, even in zero-shot learning scenarios; (2) linguistic space exploration with molecular guidance, where HME generates textual descriptions with high qualities~(+11.6\% BLEU) for molecules. These results highlight the precision of HME in handling multi-objective and cross-domain tasks, as well as its remarkable generalization capability on unseen task combinations.
HME offers a new perspective on navigating the chemical-linguistic sharing space, advancing the potential of CLMs in both fundamental research and practical applications in chemistry.

\end{abstract}
\begin{document}
\flushbottom
\maketitle

 \section*{Introduction}

In recent years, the exploration of chemical space has become a cornerstone of modern chemistry and molecular science~\cite{restrepo2022chemical,gromski2019explore}. With the scaling growth of molecular structures and their textual labels, a new paradigm is emerging: the chemical-linguistic sharing space~\cite{wozniak2018linguistic,zhou2024navigating} to enable more intuitive representations of molecular properties and interactions. To model this space, Chemical Language Models~(CLMs) have been leveraged, using sequence-based representations, including SMILES~\cite{weininger1988smiles} and SELFIES~\cite{krenn2020selfies}, to treat molecules as one-dimensional token sequences, akin to natural language~\cite{molt5,pei2023biot5,fang2023molinst}. Through masked or causal language modeling, CLMs can learn the bidirectional dependencies between molecular structures and natural language, using natural language as a medium for conditional molecular generation~\cite{elton2019deep,brown2019guacamol} and molecular comprehension~\cite{li2024empowering,tran2024mol2lang}, offering significant potential for drug discovery and virtual screening~\cite{qureshi2023ai,zongying2024taxdiff}. However, the reliance on molecular sequential representations limits the ability of CLMs to capture the full complexity and multidimensional nature of molecular structures, hindering their effectiveness in modeling complex chemical phenomena~\cite{pei2024survey_biomolecule_llm}.

Advances in molecular representation learning have led to the development of highly effective molecular structural encoders aiming at 2D graph structures~\cite{wang2022molr} and 3D coordinates~\cite{zhou2023unimol} of molecules. Integrating molecular encoders with LLMs can result in powerful CLMs capable of bridging natural language and molecular representations, unlocking new potential in molecular design and analysis~\cite{zhang2024survey_scientific_llm,zhang2024survey_sci_llm_zhangyu}.
Previous works have proposed aligning methods to effectively map the feature space of molecular encoders with the semantic space of CLMs through contrastive learning~\cite{moleculestm} or generative pretraining~\cite{cao2023instructmol,3dmolm}. However, a long-standing challenge is the inherent biases present in molecular representations from different perspectives, including 1D SMILES, 2D graph structures, and 3D conformations, which hinder CLMs from fully capturing comprehensive molecular information. Specifically, (1) SMILES sequences, generated via depth-first traversal, disrupt atomic permutation invariance, making small molecular changes cause disproportionate sequence variations~\cite{skinnider2024invalidsmiles}, introducing challenges in molecular design; (2) 2D molecular graphs, while excelling at capturing topology, face limitations such as over-smoothing and over-squashing in GNNs~\cite{chen2020oversmoothing}, and Graph Transformers may overly emphasize global interactions at the expense of local structural details~\cite{xing2024overglobal}; (3) 3D geometric models, focusing on spatial atomic arrangements and molecular conformations, struggle to differentiate molecules with similar geometries but distinct structures, recognize equivalent conformations, and maintain robustness against environmental perturbations~\cite{crippen1988conformation}. 

Some works~\cite{liu2022graphmvp,lyy-blending2d3d,lyy-unicorn2d3d} have attempted to address these biases, but they primarily focus on the inconsistencies among different molecular representations while neglecting the linguistic space. Other works, while considering the chemical-linguistic sharing space, either neglect the bias~\cite{cao2023instructmol,3dmolm} or fail to utilize LLMs to model the language space~\cite{moleculestm,feng2024bioactivity}. Moreover, existing CLMs cannot achieve effective linguistic-to-chemical exploration, e.g. molecule design, with the hindrance of molecular encoders. CLMs based on discrete sequence modeling are restricted to sequence-based representations such as SMILES. The bias of SMILES makes it challenging to meet the demands of complex drug design scenarios, such as designing molecules with specific pharmacophores~\cite{khedkar2007pharmacophore}.

To address the problems mentioned above, we propose \textbf{HME}, \textbf{H}eterogeneous \textbf{M}olecular \textbf{E}ncoding for unbiased molecular encoding to improve the bidirectional mapping within the chemical-linguistic sharing space of CLMs. Our HME is a streamlined framework that is suitable for current LLMs. 
Specifically, we encode the molecule input from sequential and geometric aspects. (1) Sequential encoding: we utilize the SMILES embedding within LLMs as the sequential molecular modeling to maximize the integration of existing chemical domain knowledge in LLMs. (2) Geometric encoding: we adopt the 2D and 3D molecule encoders as the geometric encoding for HME. We apply learnable query tokens as feature compression to address the overly sparse issue caused by dimensionality expansion and to align molecules of varying sizes. We also develop a fusion module to preemptively integrate 2D and 3D features, thereby correcting potential errors in the geometric features~\cite{van1982error3D,tang2024cycle3d,labute2005error3D}. 

Moreover, we introduce a novel molecular representation, molecular fragment sequence, into CLMs, which incorporates molecular geometric information and functional group information while presented in a string format. Specifically, we employ a subgraph mining algorithm on 2D molecular graphs to construct a fixed-size fragment vocabulary. We expand the CLM's native vocabulary and embedding layer accordingly, enabling the model to tokenize both text and fragment uniformly. Our experiments reveal that molecular fragments supplement structured information and facilitate reasoning in LLMs during generation and comprehension tasks by serving as a Chain of Thought~(CoT)~\cite{wei2022cot,chu2023survey,mu2024embodiedgpt}. The presence of molecular fragments also enables CLMs to facilitate the design of molecules with specific fragments, offering new insights into drug design.


To empirically validate the effectiveness and transferability of the HME in unified molecular tasks in the chemical-linguistic space, we integrate the HME with different LLM backbones and evaluate their performance on both molecular comprehension and generation tasks.
From the molecular comprehension aspect, our HME achieves substantial performance on molecular captioning and molecular question-answering on molecular properties and descriptions, demonstrating the molecule-text modeling ability in the chemical-linguistic sharing space. 
From the molecular generation aspect, we first constructed a novel large-scale \textbf{M}ulti-\textbf{C}onditional \textbf{Mo}lecular \textbf{D}esign~(MCMoD) dataset, which includes 1 million molecules along with their corresponding textual descriptions, molecular fragments, and chemical property control signals. 
Experiments demonstrate that our HME can reliably perform molecule design in the chemical-linguistic space with the navigation of molecular fragment targets and chemical property targets including LogP, QED, etc, which inextricably correlate with drug discovery.
Additionally, our HME demonstrates remarkable generalization and practicality. In zero-shot scenarios, HME excels at completing the multi-objective molecular inverse design task, achieving a high success rate of 79.4\%.

\section*{Results}
\subsection*{Evaluation of Molecular Encoding Bias}
Directly analyzing features derived from different molecular encodings presents challenges, such as inconsistent dimensionality and significant differences in value distributions~\cite{koppen2000cursedimension}. To address these issues, we analyze the correlation of similarity matrices computed for a given molecular set under different encodings. Specifically, if a pair of molecules exhibits high similarity under one encoding but low similarity under another, it indicates a bias between the two encodings. Using this approach, we transformed the complex features into standardized similarity values ranging from 0-1 via similarity matrices, enabling comparability across different encoding methods.

Specifically, we selected 1D, 2D, and 3D molecular encoders~\cite{honda2019smiles,moleculestm,zhou2023unimol}, and Morgan fingerprints which serves as the reference and substructure indicator~\cite{morgan1965morgan}. 
Since the features generated by molecular encoders are continuous numerical values, we use cosine similarity to measure the similarity between molecular pairs $[\mathcal{M}_i$, $\mathcal{M}_j]$ from the molecule set $\mathcal{M}$. For Morgan fingerprints, which are discrete values, we employ Tanimoto similarity instead~\cite{bajusz2015tanimoto}. 
We also calculate the similarity of features from our unbiased molecular encoder, HME, with other encoding strategies. Consequently, we obtained five similarity matrices for $\mathcal{M}$ under the five encodings. Finally, we analyzed the Pearson correlation coefficients among these five matrices.

The coefficient heatmap in Supplementary~Fig.~1 shows that: 
(1) Representational biases exist within the 1D, 2D, and 3D encoders, as indicated by correlation coefficients all below 0.15. This highlights the necessity of an unbiased and unified encoder. 
(2) Using Morgan fingerprints as the reference, HME exhibits the highest correlation, outperforming the molecular encoder from 1D, 2D, and 3D perspectives. This suggests that HME effectively captures molecular substructure information.

\subsection*{Construction of our Heterogeneous Molecular Encoding Structure}
\begin{figure*}
  \centering
  \includegraphics[width=1.0\textwidth]{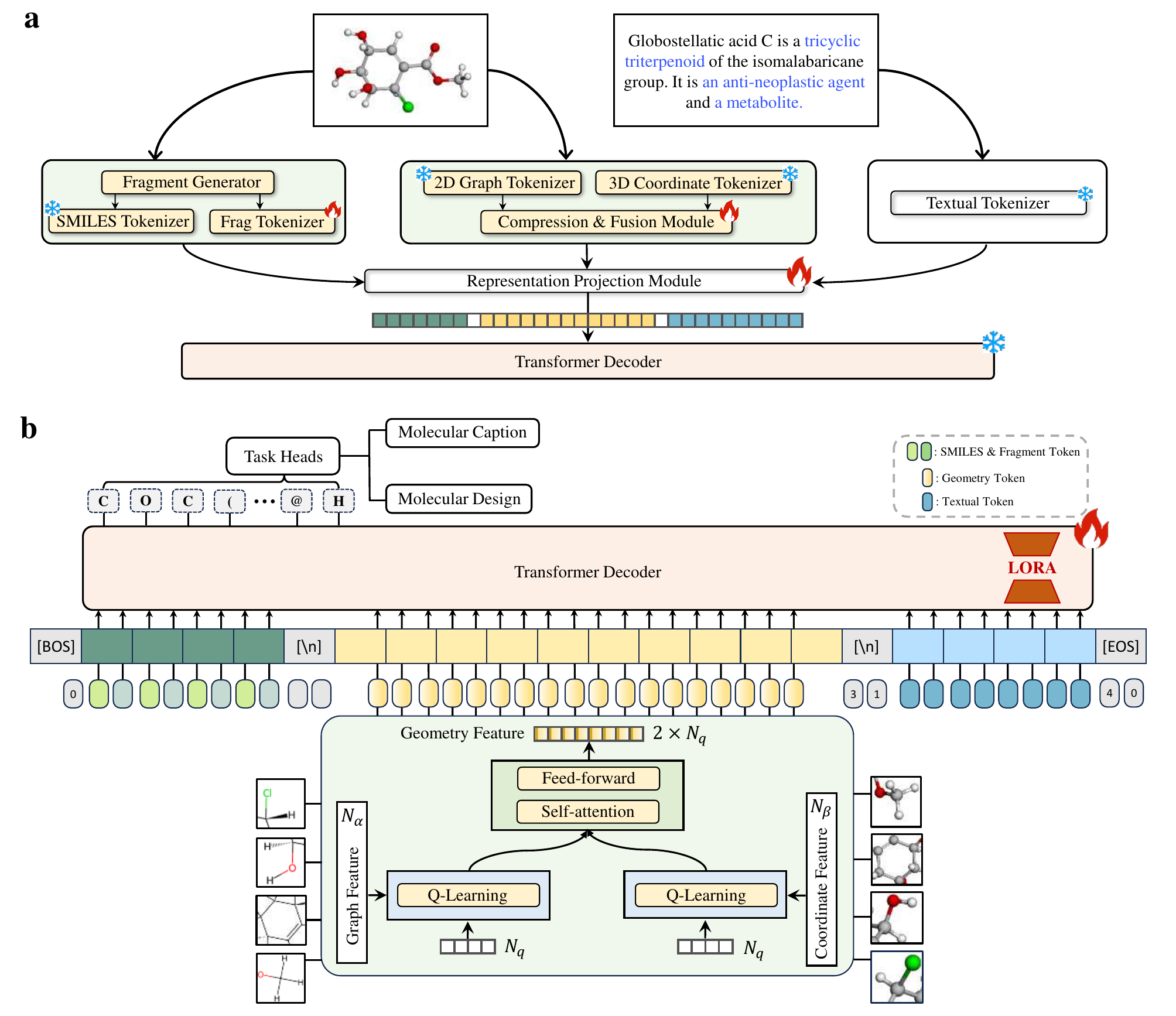}
  \caption{
  \textbf{The Framework of our HME.} (a) The alignment stage. We project features from different molecular encoders~(i.e., tokenizers) into a unified space and align them with the textual space. (b) The task-specific supervised fine-tuning stage. Based on low-rank decomposition, we perform autoregressive generative modeling on a general transformer decoder, which can generate text, molecules, and fragments. 
  }
  \label{fig:model-archi}
\end{figure*}

Fig.~\ref{fig:model-archi} illustrates the overall workflow of our HME framework. HME incorporates features of the same molecule from different perspectives to achieve deep and unbiased knowledge. As shown in Fig.~\ref{fig:model-archi}(a), HME derives encoding from four molecular encoders: (1) the 1D SMILES Encoder, which embeds the SMILES string of the molecule into the text feature space, maximizing the use of the pretrained knowledge within the CLM; (2) the 2D Graph Encoder, which encodes the molecular connectivity graph to provide geometric information; (3) the 3D Coordinate Encoder, which encodes the Euclidean coordinates of each atom, supplying more fine-grained geometric information; and (4) the Molecular Fragment Encoder, which segments the full molecule into fragments to provide significant substructure information, where fragments serve as new tokens that expand the CLM's vocabulary, offering molecular substructure information explicitly.

Fig.~\ref{fig:model-archi}(b) provides further details of HME. Using Q-learning~\cite{li2024decoupled}, we perform dimensionality augmentation on molecular geometric features while standardizing feature length, which prevents information loss and addresses the inefficiency in learning caused by varying atom counts. We employ a self-attention-based fusion layer to obtain more robust geometric features that incorporate both 2D and 3D information. We align the geometric features, 1D features, and fragment features with textual features through the projection module in Fig.~\ref{fig:model-archi}(a), and then input them together into a transformer decoder for unified autoregressive modeling. This approach enables both molecular comprehension and molecular generation to be framed as next-token probability modeling.  With fragment tokens, the Chain of Thought~(CoT)~\cite{wei2022cot} reasoning from general LLMs can be applied to molecular design, enabling HME to design essential fragment sequences before designing the complete molecule. Moreover, by using fragment sequences as a novel control condition, HME expands the scope of molecular inverse design from targeting molecular properties to targeting desired fragments.

\subsection*{MCMoD: A Large Scale Dataset for Multi-Conditional Molecular Design}
Our HME is capable of bidirectionally modeling the chemical-linguistic space, meaning it can seamlessly unify both molecular comprehension and molecular generation. This dual capability makes it essential to systematically evaluate its performance across these two major domains. In the molecular comprehension field, there are well-established and comprehensive benchmarks for evaluating the performance of CLMs~\cite{3dmolm,wu2018moleculenet,luo2023airbiomedgpt}. In the conditional molecular generation domain, benchmarks exist for evaluating molecular generation with description-based control conditions~\cite{edwards2021text2molchebi20,fang2023molinst}; however, these benchmarks are limited by their single-condition focus and small dataset size, which hampers a comprehensive assessment of CLMs' conditional molecular generation capabilities. To address these limitations, we constructed the Multi-Condition Molecular Design dataset~(MCMoD).

MCMoD includes diverse control conditions such as textual descriptions, property values, and molecular fragments, and to the best of our knowledge, it is the first dataset to introduce nearly a thousand kinds of substructures into the molecular design task. MCMoD comprises over one million molecules, covering a wide range of molecule types such as synthetic molecules, natural products, and protein ligands, differentiated by prompts. It supports multi-objective joint control, such as using both property values and specific fragments as conditions. Additionally, MCMoD creatively leverages fragment sequences as a bridge between text and molecular modalities to explore the application of the CoT paradigm in CLMs.

Overall, the MCMoD dataset we propose not only enables the systematic evaluation of HME and other CLMs in conditional molecular generation but also promotes practical applications in fields like drug development and chemical engineering through its diverse control conditions~\cite{kang2018conditionalmolgen}. The source data for MCMoD were obtained from datasets and databases such as PubChem~\cite{kim2016pubchem}, ZINC~\cite{irwin2005zinc}, ChEBI~\cite{degtyarenko2007chebi}, COCONUT~\cite{sorokina2021coconut}, and DTP~\cite{monga2002dtp}. We used RDKit to canonicalize the molecules and calculate commonly used property values, QuickVina-2~\cite{alhossary2015quickvina2} and SMINA~\cite{koes2013smina} to compute ligand binding affinities, and our fragment generator to obtain fragment sequences. Further details are provided in Supplementary~Section~2.

\begin{figure*}[t]
  \centering
  \includegraphics[width=0.98\textwidth]{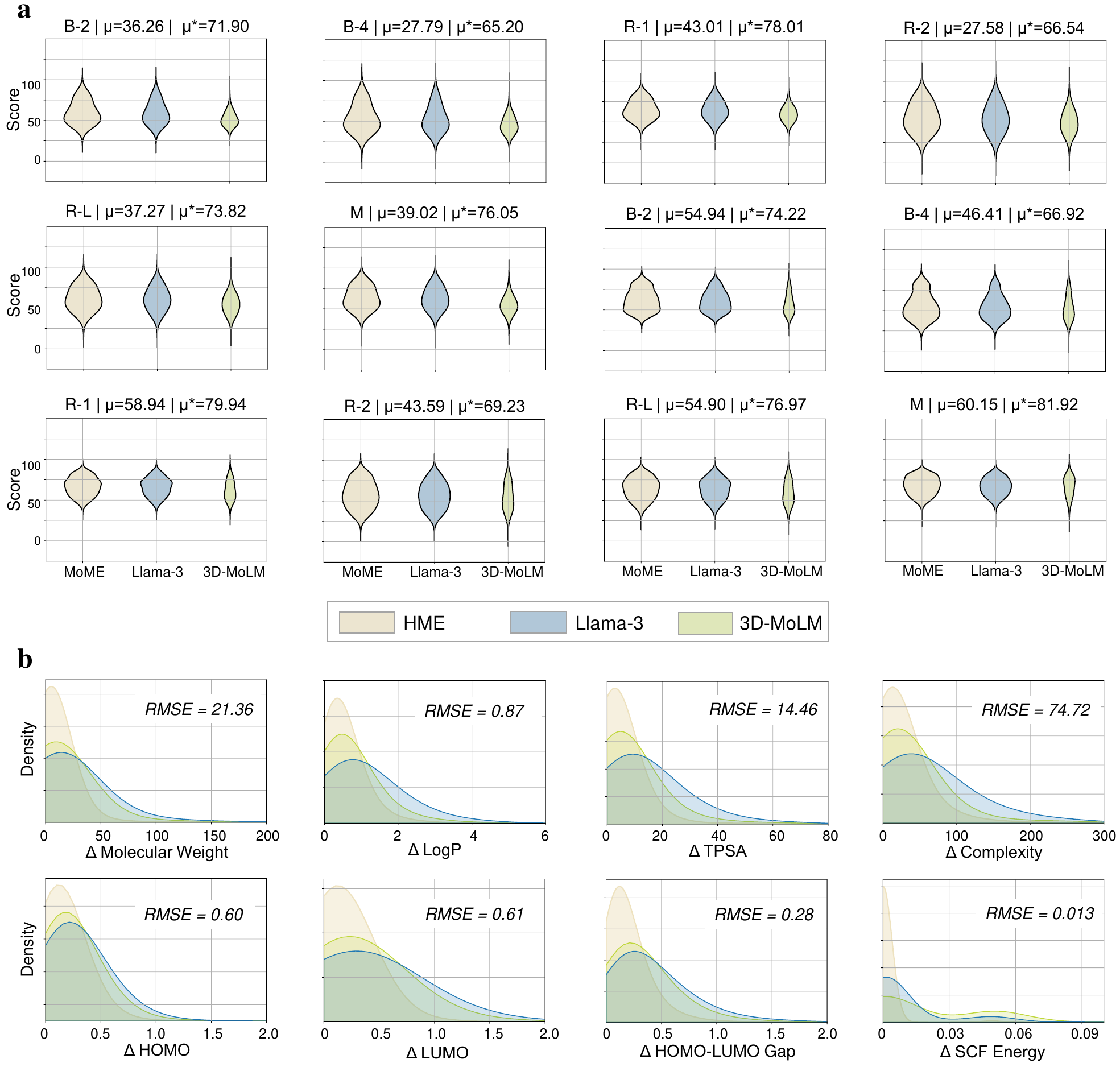}
  \caption{\textbf{Experimental Analysis for Molecular Comprehension.} \textbf{a} Captioning Task and General QA Task: The violin plot of the similarity scores between generated texts and reference texts from HME and baselines. The area size is positively correlated with the proportion of high-quality texts. B-2 means the metric BLEU-2, R-1 means the metric ROUGE-1, and M means the metric Meteor. $\mu$ means the average score of all texts from HME and $\mu^*$ means the average score of high-quality texts from HME. \textbf{b} Property QA Task: The distribution of the absolute error between the prediction values and ground-truth values from HME and baselines. The error of HME is closer to zero, demonstrating its strong capability in molecular property prediction. The value of Root Mean Square Error~(RMSE) is also reported.}
  \label{fig:2}
\end{figure*}

\begin{figure*}[t]
  \centering
  \includegraphics[width=0.98\textwidth]{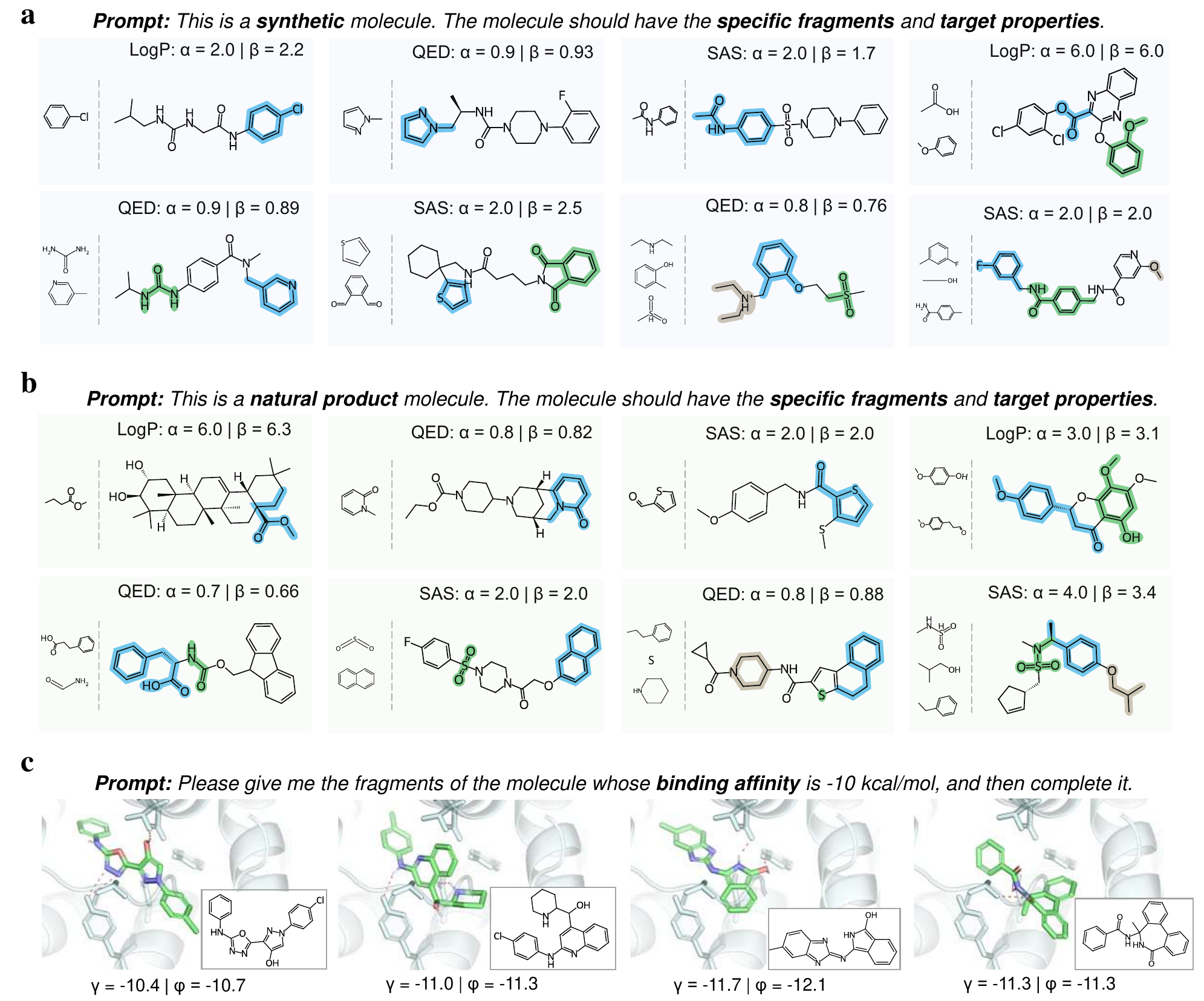}
  \caption{\textbf{Visualization for Conditional Molecular Generation of HME.} \textbf{a} We visualize eight synthetic molecules designed by HME under the joint control of property values and specific fragments. $\alpha$ denotes the target property value while $\beta$ denotes the actual value calculated by RDKit. To the left of the dotted line are the specified fragments, which are also highlighted in the generated molecules. \textbf{b} Similarly, we visualize eight natural-product-like molecules designed by HME. These cases validate that HME can design molecules with the property values and fragments we specify. \textbf{c} We visualize protein ligands with potential high affinity designed by HME. $\gamma$ and $\phi$ denote the predicted docking scores by QuickVINA2 and SMINA.}
  \label{fig:3}
\end{figure*}

\begin{figure*}[t]
  \centering
  \includegraphics[width=0.93\textwidth]{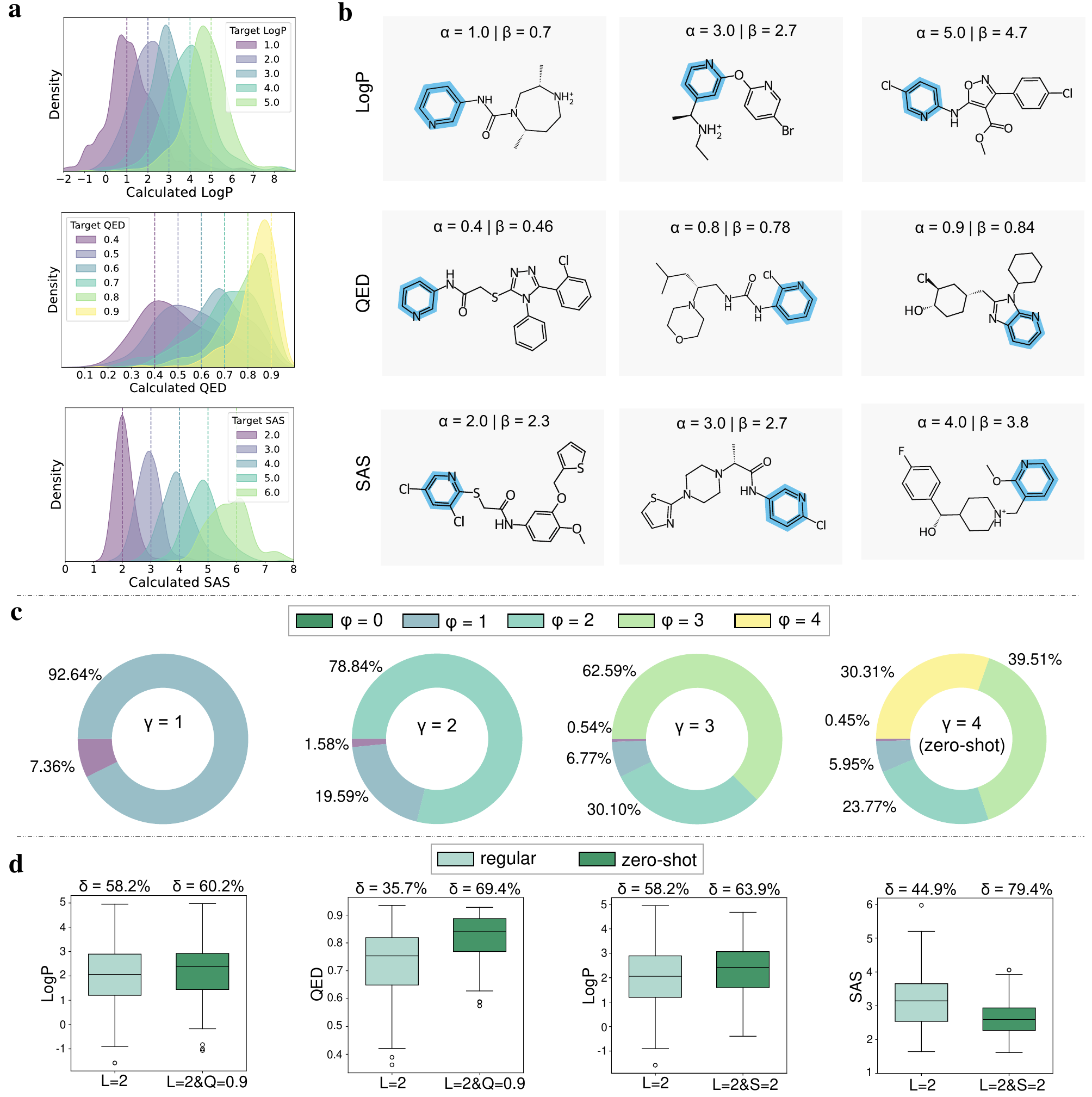}
  \vspace{-0.2cm}
  \caption{\textbf{Experimental Analysis for Conditional Molecular Generation.} \textbf{a} The distribution of the actual property values calculated by RDKit of the generated molecules under the control of the target property. Our model can effectively follow property controls. \textbf{b} Nine examples to demonstrate the effectiveness of property control. With the Pyridine as the anchor fragment condition, we use different target property types and values as property conditions. $\alpha$ denotes the target property value while $\beta$ denotes the actual value. \textbf{c} Statistics on the effectiveness of the fragment control condition, where $\gamma$ represents the specified fragment number and $\phi$ denotes the number of fragments appearing in the generated molecule. \textbf{d} The generalization ability of HME in conditional molecular generation. L, Q, and S represent the abbreviations for LogP, QED, and SAS, respectively. $\delta$ represents the proportion of molecules whose actual property values fall within the target property value range of $\pm 1$ or $\pm 0.1$. The molecules generated by HME in a zero-shot manner align well with the desired properties.}
  \label{fig:4}
\end{figure*}

\subsection*{Evaluation on Molecular Comprehension Tasks}
Molecule comprehension aims to model the mapping from chemical space to linguistic space, that is, to capture chemical structural information and express it in human-understandable natural language. This capability is crucial for bridging the gap between computational chemistry and human intuition, facilitating knowledge transfer, and enabling more effective drug discovery and documentation~\cite{2022zeng-kvplm}. However, existing approaches either lack comprehensive utilization of multi-perspective molecular features~\cite{molt5,momu} or fail to leverage LLMs --- powerful interactive tools containing rich domain knowledge~\cite{moleculestm}. Our proposed HME, built upon advanced LLMs, unifies the molecular captioning task, the general QA task, and the property QA task (detailed in respective sections) within a natural language dialogue framework. Notably, HME explicitly incorporates molecular information across three primary representations and fragment characteristics. This multi-modal approach enables a more holistic understanding and interpretation of molecular structures.

We used the 3D-MoIT dataset~\cite{3dmolm} specifically designed for CLMs as the molecular comprehension benchmark. Details of the dataset can be found in Supplementary~Section~2. We performed additional processing on the original dataset, such as obtaining graph structures, coordinates, and fragment sequences. We selected a broad range of baselines: MolT5~\cite{molt5}, which focuses on translating between SMILES and text; MoMu~\cite{momu}, which leverages both SMILES and graphs to understand molecules; 3D-MoLM~\cite{3dmolm}, which extensively utilizes molecular coordinate information; and Llama-2-7B~\cite{touvron2023llama2} and Llama-3-8B~\cite{llama3modelcard}, which are advanced LLMs rich in chemical knowledge~\cite{guo2023canllmchem}. To ensure fairness, all baselines were trained on the same dataset as our HME.

\begin{table*}[t]
\caption{\textbf{Performance Comparison for Molecular Comprehension.} We propose the experimental results of our HME and baselines. Our HME surpasses baselines under all metrics. * denotes the transformer decoder is initialized from Llama-2-7B rather than Llama-3-8B. We show one case of molecular general QA.}  \label{tab:1}
\begin{subtable}[t]{\textwidth}
\centering
\footnotesize
\begin{tabularx}{\textwidth}{ll>{\centering\arraybackslash}X>{\centering\arraybackslash}X>{\centering\arraybackslash}X>{\centering\arraybackslash}X>{\centering\arraybackslash}X>{\centering\arraybackslash}X}
\hline
\rowcolor{tableHead} \textbf{Task} & \textbf{Model} &\textbf{ BLEU-2$\uparrow$} & \textbf{BLEU-4$\uparrow$} & \textbf{ROUGE-1$\uparrow$} & \textbf{ROUGE-2$\uparrow$} & \textbf{ROUGE-L$\uparrow$} & \textbf{METEOR$\uparrow$} \\
\hline
Captioning & MolT5-Large & 25.87 & 17.28 & 34.07 & 16.42 & 23.41 & 28.04 \\
\hline
\rowcolor{tableContent} & MoMu-Large & 26.34 & 18.01 & 34.75 & 16.86 & 24.76 & 28.73 \\
\hline
 & 3D-MoLM & 30.32 & 22.52 & 36.84 & 22.32 & 31.23 & 33.06 \\
\hline
\rowcolor{tableContent} & Llama-3-8B &32.50&23.87&38.73&23.28&33.26&34.42 \\
 \hline
  & HME* & 34.29&25.44&40.95&25.12&35.28&36.67 \\
\hline
\rowcolor{tableContent} & HME & \textbf{36.26}&\textbf{27.79}&\textbf{43.01}&\textbf{27.58}&\textbf{37.27}&\textbf{39.02} \\
\hline
General QA & Llama-2-7B  & 28.15 &  23.24 &  35.14 &  22.08  & 30.41 &  46.87 \\
\hline
\rowcolor{tableContent} & 2D-MoLM & 30.84  & 25.09 &  38.46 &  24.22  & 33.04  & 50.92 \\
\hline
 & 3D-MoLM & 32.00& 26.13& 40.13& 25.55& 34.64& 52.15 \\
\hline
 \rowcolor{tableContent} & HME* & 52.24 & 43.64 & 55.71 & 40.08 & 51.56 & 57.44 \\
\hline
 & HME & \textbf{54.94} & \textbf{46.41} & \textbf{58.94} & \textbf{43.59} & \textbf{54.90} & \textbf{60.15} \\
\hline
\end{tabularx}
\end{subtable}
\centering

\vspace{0.2cm}

\begin{subtable}[t]{0.8\textwidth}
\footnotesize
\centering
\begin{tabular}{clll}
\toprule
     \textbf{Molecule} & \textbf{Question} & \textbf{Ground truth} & \textbf{Our answer} \\
\midrule
    \begin{tabular}[b]{c}
    \vspace{-1.8em}  
    \includegraphics[width=0.2\textwidth]{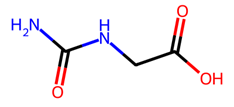} 
    \end{tabular} & 
    \begin{tabularx}{0.15\textwidth}{@{}X@{}}
    What are the \linebreak physical properties 
    of this molecule?
    \end{tabularx} & 
    \begin{tabularx}{0.27\textwidth}{@{}X@{}}
    It is a \textcolor{blue}{white crystalline solid} with a
    molecular weight of \textcolor{blue}{132.115 g/mol}.  It 
    has a melting point of around \textcolor{blue}{237-238$^{\circ}$C} 
    and is \textcolor{blue}{sparingly soluble in water}.
    \end{tabularx} & 
    \begin{tabularx}{0.27\textwidth}{@{}X@{}}
    It is a \textcolor{blue}{white crystalline solid} with a 
    molecular weight of \textcolor{blue}{130.11 g/mol}. It
    has a melting point of approximately 
    \textcolor{blue}{235-240$^{\circ}$C} and is \textcolor{blue}{sparingly soluble 
    in water}.
    \end{tabularx} \\
\bottomrule
\end{tabular}
\end{subtable}

\end{table*}

\textbf{Molecular Captioning.}
Analogous to image captioning, molecular captioning aims to generate descriptive text for molecules, a crucial task that bridges the semantic gap between chemical structures and human understanding, enabling more informed decision-making in drug discovery~\cite{molt5}.
We use common metrics from the text generation domain, such as BLEU-2, BLEU-4, ROUGE-1, ROUGE-2, ROUGE-L and METEOR, to measure the similarity between the generated text and the reference text. Specifically, BLEU focuses on precision, ROUGE emphasizes recall, and METEOR balances them.

As shown in Table.\ref{tab:1}, HME surpasses baselines across every metric. We observe that MolT5, which relies solely on 1D SMILES representations, performs worse than the three methods that incorporate molecular geometric information, despite being pretrained on the largest molecular dataset (100M molecules). This result suggests that 1D molecular information alone is insufficient for a comprehensive understanding of molecular structures. Although Llama-3-8B performs slightly better than 3D-MoLM, this is largely due to its common knowledge from pretraining on large corpora. 
HME outperforms Llama-3-8B by more than 4 percentage points across all metrics, highlighting the critical role of molecular geometric information.

In Fig.~\ref{fig:2}(a), we focus on high-quality (e.g., BLEU-2 $>$ 0.5) generated text to better analyze the model's performance. The violin plot of the scores for high-quality text is presented, where a larger area of the violin indicates a greater number of high-quality text instances. As observed from the first six plots, HME generates significantly more high-quality text compared to the baselines, with notably high average scores (e.g., $\mu$* = 71.90). This demonstrates that HME performs precise molecular captioning.

\textbf{Molecular General QA.}
Recently, LLMs such as ChatGPT have demonstrated the ability to answer a wide range of human queries~\cite{ouyang2022instructgpt}. Naturally, this has led to an emerging demand for asking questions about molecules, in order to gain insights into their physicochemical properties, functions, origins, and other aspects. As a result, the Molecular General QA task has arisen to meet this need, enabling users to interact with models in a way that provides comprehensive and accessible molecular knowledge. The same metrics used in the molecule captioning task are used to evaluate the correctness of the generated answers. 
HME also outperforms the baselines in this task, as shown in Table.~\ref{tab:1}. It achieves significantly higher improvements across all metrics compared to the baselines; for instance, HME scores 43.59 on the ROUGE-2 metric, compared to 25.55 by 3D-MoLM. It may be due to: (1) HME captures substantially more molecular information than the baselines; and (2) the Q-Former-based 3D-MoLM excessively compresses molecular features, leading to information loss, whereas our HME shortens feature length while simultaneously expanding feature dimensions, ensuring both molecular and textual information are preserved and effectively utilized by the downstream decoder. This highlights the rationality of our HME architecture.

We also draw the violin plots to visualize the score distribution of the high-quality generated text in Fig~.\ref{fig:2}(a). 
As observed in the last six plots, although HME and Llama-3-8B share the same transformer decoder, HME still generates more answers of higher quality in the QA task. The concrete case in Table.~\ref{tab:1} demonstrates that HME accurately addresses critical molecular details, including color, function, melting point, and solubility, validating the effectiveness of HME.

\textbf{Molecular Property QA.}
The objective of molecular property prediction is to infer diverse property values based on molecular structures. Within our HME framework, molecular property prediction is transformed into a Molecular Property QA paradigm, enabling users to specify properties of interest through natural language queries. This approach gives rise to a more user-friendly and unified model capable of predicting multiple molecular properties simultaneously.
Our HME model can predict the property values of a given molecule, including not only conventional properties such as molecular weight, LogP, and TPSA, but also quantum chemical properties like HOMO, LUMO, HOMO-LUMO Gap, and SCF Energy. To evaluate the model's performance, we extract the predicted numerical values from the generated answers using regular expressions, and then compare them with the ground-truth values.

As shown in Fig.~\ref{fig:2}(b), we plot the distribution of the absolute difference between the predicted and actual values. We observe that 3D-MoLM outperforms Llama-3-8B, which only encodes 1D information, in 7 properties out of 8. This verifies that the inclusion of 3D information helps the model better learn molecular properties. Furthermore, HME significantly outperforms 3D-MoLM, demonstrating the effectiveness of heterogeneous encoding. Specifically, 2D graphs provide a more stable molecular representation, containing expert knowledge such as bond types and atomic connectivity. Although 3D coordinates offer detailed spatial information, they may carry potential errors and lack stability. By integrating 2D and 3D information, HME gains a robust and comprehensive understanding of molecular geometry. In Supplementary~Section~5, we present more results of the property QA task. In Supplementary~Section~4, we present the performance in protein-ligand affinity prediction, which demonstrates the practical applicability of HME.

\begin{table*}[t]
\centering
\footnotesize
\caption{\textbf{Performance Comparison for Description-based Molecular Generation.} We propose the experimental results on our HME and baselines including RNN and advanced LLMs. Taking fragment sequences as Chain of Thought~(CoT), our HME surpasses baselines under all metrics except Validity. One case is proposed for the visualization: The text highlight area and the fragment highlight area have a corresponding relationship, which reveals the interpretability and transparency of CoT.} \label{tab:2}
\begin{subtable}{\textwidth}
\begin{tabularx}{\textwidth}{l>{\centering\arraybackslash}X>{\centering\arraybackslash}X>{\centering\arraybackslash}X>{\centering\arraybackslash}X>{\centering\arraybackslash}X>{\centering\arraybackslash}X>{\centering\arraybackslash}X>{\centering\arraybackslash}X>{\centering\arraybackslash}X}

\hline
\rowcolor{tableHead} \textbf{Models} & \textbf{BLEU$\uparrow$} & \textbf{Exact$\uparrow$} & \textbf{Levenshtein$\downarrow$} & \textbf{\mbox{MACCS FTS$\uparrow$}} & \textbf{RDK FTS$\uparrow$} & \textbf{\mbox{Morgan FTS$\uparrow$}} & \textbf{FCD$\downarrow$} & \textbf{Validity$\uparrow$} \\
\hline
RNN & 0.652 & 0.005 & 38.09 & 0.591 & 0.400 & 0.362 & 4.55 & 0.542 \\
\hline
\rowcolor{tableContent} Transformer & 0.499 & 0.000 & 57.66 & 0.480 & 0.320 & 0.217 & 11.32 & 0.906 \\
\hline
MolT5-Small & 0.755 & 0.079 & 25.988 & 0.703 & 0.568 & 0.517 & 2.49 & 0.721 \\
\hline
\rowcolor{tableContent} MolT5-Base & 0.769 & 0.081 & 24.458 & 0.721 &  0.588 & 0.529 & 2.18 & 0.772 \\
\hline
MolT5-Large & {0.854} & {0.311} & {16.071} & 0.834 & 0.746 & {0.684} & 1.20 & 0.905 \\
\hline
\rowcolor{tableContent} GPT-3.5~(zero-shot) & 0.489 & 0.019 & 52.13 & 0.705 & 0.462 & 0.367 & 2.05 & 0.802 \\
\hline
 GPT-3.5~(10-shot) & 0.790 & 0.139 & 24.91 & 0.847 & 0.708 & 0.624 & 0.57 & 0.887 \\
\hline
\rowcolor{tableContent} MolXPT & - & 0.215 & - & {0.859} & {0.757} & 0.667 & 0.45 & \textbf{0.983} \\
\hline
 HME & \textbf{0.879} & \textbf{0.334} & \textbf{14.25} & \textbf{0.908} & \textbf{0.817} & \textbf{0.753} & \textbf{0.41} & {0.910} \\
\hline
\end{tabularx}
\end{subtable}

\vspace{0.2cm}

\begin{subtable}[t]{0.8\textwidth}
\footnotesize
\centering
\begin{tabular}{clll}
\toprule
     \textbf{Instruction} & \textbf{Description} & \textbf{Generated Molecule}  \\
\midrule
    \begin{tabularx}{0.3\textwidth}{@{}X@{}}
    Please give me molecular fragments based on the description. And then give me the molecule based on the fragments.
    \end{tabularx} &
    \begin{tabularx}{0.3\textwidth}{@{}X@{}}
    The molecule is a 3',5'-cyclic purine nucleotide that is 3',5'-cyclic AMP bearing \textcolor{blue}{an additional bromo substituent} at position 8 \textcolor{blue}{on the adenine ring}. It has a role as a protein kinase agonist and an antidepressant. \end{tabularx}
    &
    \begin{tabular}[b]{c}
    \vspace{-3.5em}  
    \includegraphics[width=0.3\textwidth]{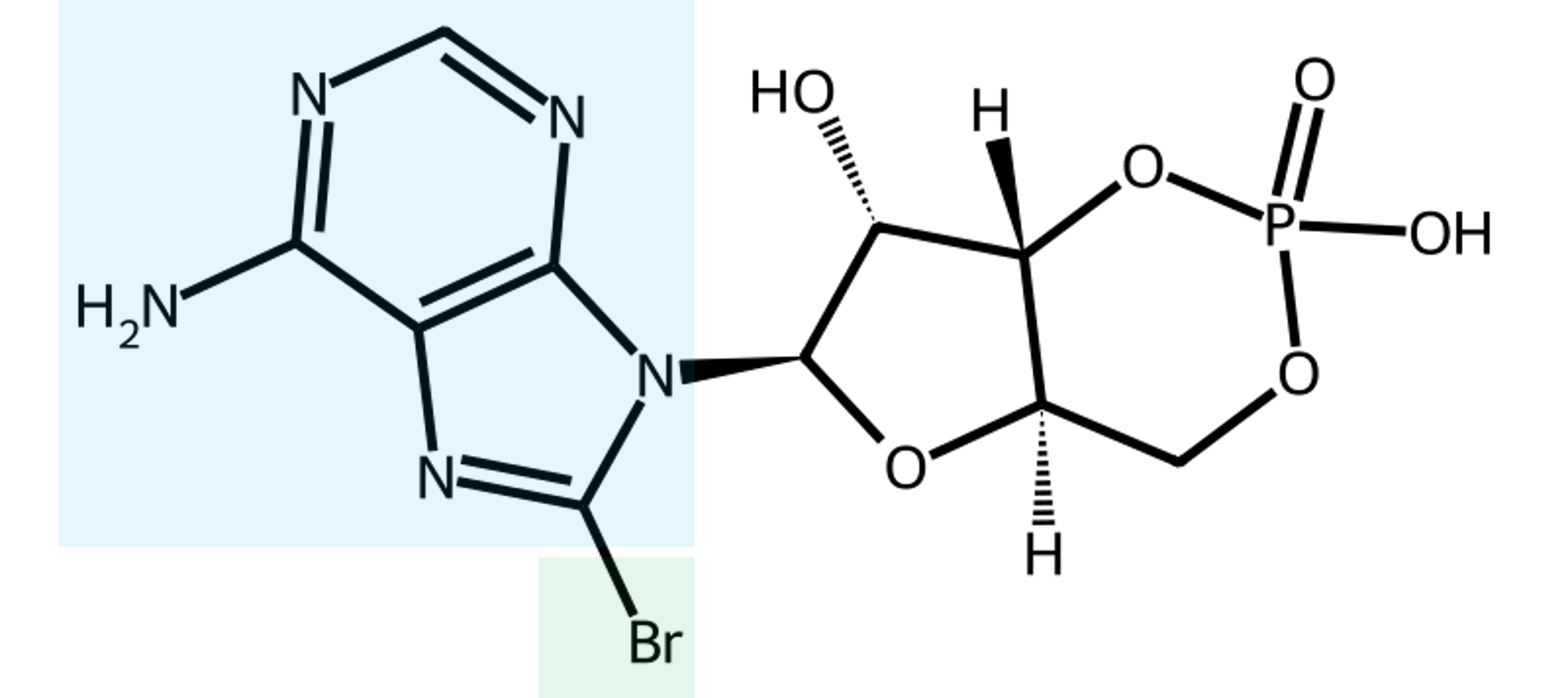} 
    \end{tabular} \\
\bottomrule
\end{tabular}
\end{subtable}

\end{table*}



\subsection*{Evaluations on Conditional Molecular Generation Tasks}

Conditional molecular generation aims to design novel molecules satisfying specified constraints, which is crucial for drug discovery and related applications as it circumvents blind exploration of the vast chemical space. Previous approaches lack a unified framework for flexible conditional control. In our work HME, we unify diverse conditional generation tasks into next-token prediction by expressing various conditions as text or fragment sequences. Leveraging the flexibility of natural language, HME demonstrates remarkable zero-shot generalization capabilities. The incorporation of molecular fragment sequences may accelerate fragment-based molecular design while bridging the text-molecule gap through CoT reasoning.
We used the MCMoD dataset as a benchmark to evaluate the performance of HME across three tasks of conditional molecular generation: multi-objective inverse molecular design, description-based molecular generation, and protein-ligand design. The baselines span a wide range, including simple RNNs~\cite{sherstinsky2020rnn}, simple transformers~\cite{vaswani2017transformer}, MolT5~\cite{molt5} models of different scales, GPT-3.5~\cite{ouyang2022instructgpt} with exceptional zero-shot/few-shot capabilities, and the generative molecular pretrained model MolXPT~\cite{liu2023molxpt}.

\textbf{Multi-Objective Molecule Reverse Design.}
The goal of multi-objective molecular inverse design is to generate molecules that satisfy multiple constraints, which is highly challenging and practical. For example, drug-like molecules often require properties such as specific solubility and optimal bioavailability. Additionally, it is desirable for designed molecules to include specific fragments, as they play a critical role in determining biological activity and target specificity~\cite{yang2024digfrag}.
Our HME model offers a novel approach to tackle this challenge. HME enables inverse molecular design under joint control of both property values and molecular fragments, represented in natural language for exceptional flexibility and generalization.
Specifically, we consider three properties---solubility (LogP), drug-likeness (QED), and synthetic accessibility (SAS)---along with 800 kinds of molecular fragments as conditions. Property values are discretized at intervals of 1 or 0.1, allowing for finer control granularity compared to prior works. The 800 fragments are sourced from the fragment vocabulary of HME, which comprises the most frequently occurring substructures in the MCMoD dataset. To the best of our knowledge, HME represents the first attempt to jointly control properties and fragments for molecular inverse design, paving the way for more precise and versatile approaches in the field.

We present 8 synthetic molecules generated by HME under different control conditions in Fig.~\ref{fig:3}(a). In the 1st and the 4th cases, we set distinctly different LogP values of 2.0 and 6.0 as property targets, along with 1 and 2 fragments as fragment targets, respectively. We observe that the generated molecules contain all the specified fragments, and their LogP values, calculated using RDKit, align closely with the specified LogP conditions. In the 2nd, 5th, and 7th cases, 1, 2, and 3 fragments are used as control conditions, respectively, with a focus on the QED property. A QED value closer to 1 indicates better drug-likeness. We observe that HME successfully generates molecules containing the specified fragments while achieving good drug-likeness (QED $>$ 0.7), demonstrating its capability to design drug candidates with fragments of interest. Similarly, in the 3rd, 6th, and 8th cases, we use SAS and fragments as joint control conditions. We observe that even when using three fragments as control conditions, HME adheres to the specified controls and generates molecules with the desired SAS values. This verifies HME's ability to generate molecules with good synthetic accessibility, which can help reduce the cost of new drug development.

As shown in Fig.~\ref{fig:3}(b), by adjusting the prompts, we directed HME to generate molecules resembling natural products, which are typically more structurally complex than synthetic molecules. Even so, HME performed remarkably well. Specifically, in the 1st case, the given fragments were not incorporated as linear chains but instead participated in forming complex ring structures, demonstrating the flexibility and versatility of fragment control. In the 7th case, the controlling fragments themselves exhibited relatively low drug-likeness, with QED values of 0.46 for pyrrolidine and 0.56 for n-butylbenzene. Nevertheless, HME identified a way to combine these fragments into a molecule with a highly drug-like structure, achieving a final QED as high as 0.88.

To further demonstrate the effectiveness of our conditional controls, we generated 3,200 molecules on the test dataset and evaluated the corresponding metrics, as shown in Fig.~\ref{fig:4}. In the first plot of Fig.~\ref{fig:4}(a), we varied the target LogP value from 1.0 to 5.0 and analyzed the distribution of the actual LogP values of the generated molecules. We observed that the peak of the distribution consistently aligns with the target LogP, and the distribution curves shift as the target LogP increases, verifying the effectiveness of LogP control. Similarly, the second and the third plots of Fig.~\ref{fig:4}(a) reflect the effectiveness of QED and SAS controls, respectively. Notably, in the second plot of Fig.~\ref{fig:4}(a), when the target QED is set to 0.9, the distribution curve becomes more concentrated, indicating that HME performs better when designing molecules with higher drug-likeness. Likewise, in the third plot of Fig.~\ref{fig:4}(a), the distribution curve is more concentrated when the target SAS is 2, suggesting that HME excels in designing molecules with better synthetic accessibility compared to molecules with poorer synthetic accessibility. This aligns well with real-world requirements, highlighting HME’s significant potential for practical applications in molecular design. 
In Fig.~\ref{fig:4}(b), we present nine cases that use the same fragment as an anchor condition alongside nine different property conditions, which showcases HME's capability for multi-objective molecular inverse design. 

The conditional fragments are effective. The first three ring charts in Fig.~\ref{fig:4}(c) illustrate the success rates of incorporating one, two, or three fragments as control conditions into the generated molecules. The complete inclusion rates were 92.64\%, 78.84\%, and 62.59\%, respectively. This indicates that despite the increasing difficulty of the task with a greater number of fragments, HME still demonstrated outstanding performance. Moreover, when considering the success of any two out of three fragments, the success rate reached as high as 92.69\%. Additionally, we observed a sharp decline in the complete exclusion rate of fragments as the number of control fragments increases, with rates of 7.36\%, 1.38\%, and 0.54\%, respectively. These results indicate HME's ability to capture the mapping relationship from fragments to complete molecules.

\textbf{Description-Based Molecular Generation.}
Opposite to molecular captioning, which generates a description from a molecule, this task aims to generate a molecule from a detailed description, as shown in Table.~\ref{tab:2}. The significance of this task lies in its potential applications, such as enabling chemists to design molecules by providing precise natural language specifications, thereby bridging human intentions and computational molecular design. While previous works focused on directly mapping textual descriptions to SMILES using CLMs, our approach, inspired by CoT reasoning and the concept of slow thinking~\cite{wei2022cot}, introduces an intermediate step for HME by first generating fragment tokens aligned with the textual description before completing the molecule. This step effectively articulates the model's reasoning process prior to producing the final output, enhancing both interpretability and transparency.
We employed multiple metrics to evaluate the similarity between the molecules generated by the models and the reference molecules, including BLEU, various molecular fingerprint similarities, and the Fréchet ChemNet Distance~(FCD)~\cite{preuer2018fcd}, which measures the distance between molecules in the chemical space.


As shown in Table.~\ref{tab:2}, HME achieves the best performance on 7 out of 8 metrics, despite baselines such as MolT5 and MolXPT being pre-trained on molecules over 1,000 times larger than ours. Compared to MolT5, HME shows a slight improvement in BLEU~(i.e., SMILES similarity) but a more substantial improvement in fingerprint similarity and FCD. We attribute this to the fragment tokens, which, although in sequence form, are derived from 2D molecular graphs and carry substantial structural information. Consequently, the CoT approach based on fragment tokens enables HME to generate molecular string representations from a perspective more closely aligned with the molecular structure.

We observe that the fragment tokens capture 2D patterns effectively. For example, the molecule, whose SMILES is ``C([C@H]([C@H]([C@@H]([C@H](CO)O)O)O)O)O”, consists of three identical ``OCCO” fragments, a combination pattern that CLMs relying solely on SMILES would struggle to identify. Furthermore, CoT enhances the transparency and interpretability of CLMs. For example, in the case of Table.~\ref{tab:2}, we colored the keywords of the textual description in blue. And we found that the fragment chain generated by HME accurately reflects these keywords, incorporating specific molecular fragments like the adenine ring and the bromine substituent. To validate the consistency of CoT, we calculated the precision and recall between the fragments included in the final molecule and those in the chain, achieving scores of 0.769 and 0.774, respectively. This supports the coherence between the reasoning process and the final output of HME.

\textbf{Affinity-Based Ligand Generation.}
Ligand design aims to generate small molecules with high binding affinity to protein pockets, a challenging task that requires implicit understanding of molecular conformations and poses. Unlike previous approaches, HME employs CoT reasoning to decompose this complex task: it first designs molecular fragments that meet affinity requirements, followed by complete molecule generation. Binding affinity is quantified by docking scores, which are rounded to negative integers ranging from -5 to -10, with lower values indicating stronger binding.
For each specified docking score, we generated 100 molecules, totaling 600 molecules, and used QuickVina-2 and SMINA~\cite{koes2013smina} to calculate their docking scores as actual values. We observed that the mean absolute error between the actual docking scores and the condition-specified docking scores was 0.46, with a Pearson correlation coefficient as high as 0.91, supporting the effectiveness of conditional control. Furthermore, since the goal of ligand design is to generate molecules with higher affinity (i.e., lower docking scores), we defined molecules with a docking score $\leq$ -10 as high affinity. We observed that when the control signal was set to -10, 20\% of the generated molecules exhibited high affinity, compared to only 8\% in the training set. In Fig.~\ref{fig:3}(c), we visualized four generated molecules with affinities exceeding those of all known molecules in the training dataset. This suggests that our method can accelerate the search through chemical space for higher-affinity ligands by specifying a condition, thus expediting virtual drug screening. 

\textbf{Molecular Generation in Zero-shot Manner.} 
Leveraging the flexibility of natural language and next-token prediction, our HME model demonstrates remarkable zero-shot capability. Specifically, in the multi-objective molecular inverse design task, HME can generate desired molecules under the control of four fragments, despite the training data containing at most three fragments. Additionally, although HME is trained with control over only a single property type, it generalizes effectively to multi-property control conditions.

As shown in Fig.~\ref{fig:4}(c) (the fourth panel), under the condition of four control fragments ($\gamma=4$), we generated 2000 molecules and evaluated how many of the control fragments were incorporated into the final molecules. We observed that 30.31\% of the molecules contained all four control fragments, which is a satisfactory result given the increased difficulty of the task with more fragments. Additionally, 69.82\% of the molecules contained at least three fragments, surpassing the 62.59\% observed for $\gamma=3$. Furthermore, 93.59\% of the molecules contained at least two fragments, significantly higher than the 78.84\% observed for $\gamma=2$. These results validate HME's strong generalization ability with respect to the number of fragments.

As illustrated in Fig.~\ref{fig:4}(c), we generated three sets of 100 molecules using the property control conditions (i) LogP = 2, (ii) LogP = 2 with QED = 0.9, and (iii) LogP = 2 with SAS = 2. The first set served as a non-zero-shot baseline, while the latter two were considered zero-shot conditions, as combinations of dual-property controls were not used during training. As shown in the first panel, under the zero-shot scenario, the actual LogP values of the generated molecules remained centered around 2, closely matching the baseline. The success rate $\delta = 60.2\%$ even slightly outperformed the baseline $\delta = 58.2\%$. A similar trend is observed in the third panel, where the distribution and success rate of LogP for the baseline and zero-shot scenarios are nearly identical. This verifies that the control of the primary property in zero-shot settings is not compromised by the addition of a second property. In the second panel, we observed that the newly added QED control condition had a significant impact, increasing $\delta$ from the baseline 35.7\% to 69.4\%, with the actual QED values of the molecules closely distributed around 0.9. Similarly, in the fourth panel, the addition of SAS control improved $\delta$ from 44.9\% to 79.4\%, with the median SAS values falling between 2 and 3. These results reflect the effectiveness of dual-property joint control in zero-shot settings.

These compelling results underscore HME's remarkable generalization capabilities in molecular reverse design, potentially accelerating the drug development pipeline and enabling more efficient exploration of the vast chemical space.

\section*{Discussion}
This work presents a unified framework called HME for modeling the chemical-linguistic space, integrating various molecular generation and comprehension tasks. We analyze feature similarities of molecular representations from different perspectives, revealing the existence of bias. We introduce a novel molecular representation, molecular fragment sequence, into CLMs and mix it with three major molecular features to help CLMs model the chemical-linguistic space more effectively and without bias. To comprehensively explore the molecular generation capability of our HME, we collect and open-source MCMoD, the first large-scale molecular design dataset with multiple conditions. The experimental validation in Fig.~\ref{fig:2} and Fig.~\ref{fig:3} demonstrates its proficient capability in both molecular comprehension and generation. Additionally, HME is the first chemical LLM that shows excellent molecular design ability with composed conditions including hundreds of fragments, which may facilitate fragment-based drug discovery. 

One limitation of our work is that compared to the large-scale datasets in the visual-language domain, datasets in this field can be further expanded to better model the chemical-linguistic space, which requires collective efforts from the entire community. Additionally, this work does not consider molecular dynamics, which is becoming increasingly important due to its crucial role in understanding protein-ligand interactions, conformational changes, and reaction mechanisms at the atomic level. In the future, we plan to extend HME to tasks related to chemical reactions and molecular dynamics. Furthermore, we intend to deeply explore pharmacophore-based molecular design and lead optimization, aiming to contribute to the field of drug design.

\section*{Methods}
Molecules can be characterized from various perspectives, including SMILES string sequences, molecular graphs, and molecular coordinates. In this work, we propose heterogeneous molecular encoding to enable CLMs to achieve a more comprehensive understanding of molecules. HME enables bidirectional modeling of the chemical-linguistic space, supporting a wide range of tasks, including molecular captioning and molecular design. In the following text, we present the architecture and training procedure of HME. The prompts used for various tasks in HME are detailed in Supplementary~Section~3.

\subsection*{Model Architecture}

\textbf{Molecular SMILES Tokenizer.} We directly leverage the tokenizer and the shallowest embedding layer of Llama-3.0-8B as the SMILES Tokenizer denoted as $\mathcal{E}_{1D}$. This approach was chosen for two reasons: LLM tokenizers can naturally process string inputs, and their extensive pretraining gives them some understanding of SMILES. As a result, we can effectively use the LLM's internal knowledge. Specifically, for an input SMILES string $\mathcal{M}_{1D}$, we first decompose it into a sequence of tokens $\{\mathbf{t}_1, \mathbf{t}_2, ..., \mathbf{t}_n\}$, where each token $t_i$ is represented as a one-hot vector $\mathbf{t}_i \in \{0,1\}^{|\mathcal{V}|}$ in the vocabulary space $\mathcal{V}$. Subsequently, $\mathcal{E}_{1D}$ transforms each token embedding through a series of learnable matrices: 
\begin{equation}
\mathbf{v}_i = \mathbf{W}_E\mathbf{t}_i + \mathbf{p}_i, \quad i \in {1,...,n},
\end{equation}
where $\mathbf{W}_E \in \mathbb{R}^{d \times |\mathcal{V}|}$ is the token embedding matrix, $\mathbf{p}_i \in \mathbb{R}^d$ denotes the positional embedding for position $i$.

\textbf{Molecular Fragment Generator.} We introduce the principle subgraph mining algorithm to construct a vocabulary of different fragments. Given an initial vocabulary consisting of atomic elements ($n_{atom}$ unique atoms), we iteratively construct the complete vocabulary of size $n$. The algorithm performs $n-n_{atom}$ merging operations. For each iteration, consider a molecular fragment $\mathcal{F}=(\mathbf{V}, \mathbf{E})$, where $\mathbf{V}$ and $\mathbf{E}$ denote its vertex and edge sets respectively. The neighborhood of $\mathcal{F}$ is defined as:
\begin{equation}
\mathcal{N}(\mathcal{F}) := \{\mathcal{F}' = (\mathbf{V}', \mathbf{E}') \mid \exists v \in \mathbf{V}, v' \in \mathbf{V}', \delta(v, v') = 1\},
\end{equation}
where $\delta(\cdot,\cdot)$ measures the topological distance between vertices. For each neighboring pair $(\mathcal{F}, \mathcal{F}')$, we compute their union $\mathcal{F} \cup \mathcal{F}'$ and maintain a frequency counter $f(\cdot)$. The fragment with maximum frequency is selected until the vocabulary reaches the target size $n$:
\begin{equation}
\mathcal{F}^* = \textrm{argmax}_{\mathcal{F} \cup \mathcal{F}'} f(\mathcal{F} \cup \mathcal{F}'), \quad \forall \mathcal{F}' \in \mathcal{N}(\mathcal{F}).
\end{equation}


After constructing the fragment vocabulary, we formalize the molecular decomposition process. Given a molecule and its initial atomic decomposition $\mathcal{F}_i (i=1,...,k)$, we iteratively merge adjacent fragments. At each step, we consider the set of all possible merging candidates $\mathcal{C}$, select the optimal merge operation $\mathcal{F}^*$:
\begin{equation}
\begin{aligned}
\mathcal{C} = \{(\mathcal{F}_i \cup \mathcal{F}_j) \mid \mathcal{F}_j \in \mathcal{N}(\mathcal{F}_i), i,j \in {1,\ldots,k}\} , \quad
\mathcal{F}^* = \textrm{argmax}_{\mathcal{F} \in \mathcal{C} \cap \mathcal{V}} f(\mathcal{F}),
\end{aligned}
\end{equation}
where $f(\cdot)$ denotes the frequency count from vocabulary construction, and $\mathcal{C} \cap \mathcal{V}$ ensures the merged fragment exists in vocabulary $\mathcal{V}$. This process continues until no valid merging operations remain.

\textbf{Molecular Fragment Tokenizer.} We augment the original vocabulary of Llama-3 with our constructed fragment vocabulary. Specifically, given the base vocabulary $\mathcal{V}_{base}$ and fragment vocabulary $\mathcal{V}_{frag}$, we construct the augmented vocabulary space:
\begin{equation}
\mathcal{V}_{aug} = \mathcal{V}_{base} \cup \mathcal{V}_{frag}, \quad |\mathcal{V}_{aug}| = |\mathcal{V}_{base}| + |\mathcal{V}_{frag}|.
\end{equation}

The embedding matrix is correspondingly expanded from $\mathbf{W}_{base} \in \mathbb{R}^{d \times |\mathcal{V}_{base}|}$ to $\mathbf{W}_{aug} \in \mathbb{R}^{d \times |\mathcal{V}_{aug}|}$. To ensure stable training dynamics, we initialize the newly introduced parameters using mean initialization with Gaussian noise:
\begin{equation}
\mathbf{W}_{aug}[i,j] = \begin{cases}
\mathbf{W}_{base}[i,j], & \text{if } j \leq |\mathcal{V}_{base}| \\
\mu + \sigma \cdot \xi_{ij}, & \text{if } j > |\mathcal{V}_{base}|,
\end{cases}
\end{equation}
where $\mu = \mathbb{E}[\mathbf{W}_{base}]$, $\sigma = \sqrt{\text{Var}[\mathbf{W}_{base}]}$, and $\xi_{ij} \sim \mathcal{N}(0,1)$ ensures statistical consistency with the pretrained embeddings.

\textbf{Molecular Graph Tokenizer.} For an input molecular graph $\mathcal{G} = (\mathbf{V}, \mathbf{E})$, where $\mathbf{V} \in \mathbb{R}^{n \times d_v}$ represents the feature matrix of nodes (with $d_v$ being the dimensionality of node features) and $\mathbf{E} \in \mathbb{R}^{m \times d_e}$ denotes the feature matrix of edges (with $d_e$ being the dimensionality of edge features), we utilize a pretrained Graph Isomorphism Network (GIN) as the encoder:
\begin{equation}
\mathbf{h}_v^{(k)} = \text{MLP}^{(k)}\left((1 + \epsilon) \cdot \mathbf{h}_v^{(k-1)} + \sum_{u \in \mathcal{N}(v)} \mathbf{h}_u^{(k-1)}\right),
\end{equation}
where $\mathbf{h}_v^{(k)}$ is the representation of node $v$ at layer $k$, $\epsilon$ is a learnable parameter, and $\mathcal{N}(v)$ denotes the set of neighbors of node $v$. The final node representations after $K$ layers serve as the 2D features of the molecule.

\textbf{Molecular Coordinate Tokenizer.} We use a pretrained transformer-based encoder to take atom coordinates as input directly. The pair-level representations are introduced into the encoder, initialized via spatial positional encodings. These representations, denoted as $\mathbf{q}_{ij}^l$ for an atom pair ($i$,$j$) at layer $l$, are updated using atom-to-pair communication based on the Query-Key product, and the pair-to-atom communication mechanism are used to incorporate 3D information into atom representations, based on bias-involved attention:
\begin{equation}
\begin{aligned}
    \mathbf{q}_{ij}^{l+1} = \mathbf{q}_{ij}^l +\{ \frac{ \mathbf{Q}_i^{l} (\mathbf{K}_j^{l})^T }{\sqrt{d}} \}, \quad
    \text{Attention}(\mathbf{Q}_i^{l}, \mathbf{K}_j^{l}, \mathbf{V}_j^{l}) = \text{softmax}\left( \frac{ \mathbf{Q}_i^{l} (\mathbf{K}_j^{l})^T }{\sqrt{d}} + \mathbf{q}_{ij}^{l-1} \right) \mathbf{V}_j^{l},
\end{aligned}
\end{equation}
where $\mathbf{Q}_i^{l}, \mathbf{K}_i^{l}, \mathbf{V}_i^{l}$ represent the Query and Key matrices of atoms $i$. The node representations of the final layer serve as the 3D features of the molecule.

\textbf{Compression, Fusion and Projector Modules.} 
Through Q-learning, we process variable-length 2D and 3D features for efficient model learning. Specifically, Q-learning compresses the information contained in 2D features and 3D features into $n$ query tokens $\textbf{Q}=\{\textbf{q}_1,...,\textbf{q}_n\}$ based on the cross-attention mechanism:
\begin{equation}
\begin{aligned}
    \textbf{U}&=\textrm{Attention}(\textbf{Q}, \textbf{W}_{up}\textbf{U}, \textbf{W}_{up}\textbf{U}),
\end{aligned}
\end{equation}
where $\textbf{U}$ is the output from the 2D or 3D Tokenizer, and $\textbf{W}_{up}$ is used to upsample the features.
We designed a fusion module to integrate 2D information $\textbf{U}_{2D}$ and 3D information $\textbf{U}_{3D}$ at a fine-grained level to obtain robust geometry information $\textbf{U}_{geo}$ with self-attention:
\begin{equation}
   \textbf{U}_{geo} = \textrm{Attention}(\textrm{Concat}(\textbf{U}_{2D}, \textbf{U}_{3D}), *, *),
\end{equation}
where $\textrm{Concat}$ means concatenating the two along the sequence length. Finally, the features obtained from different tokenizers are directly concatenated along the token length to ensure no information is lost.

\textbf{AutoRegressive Decoder.} We stack 32 transformer decoder layers as a universal decoder, which is used to extract the high-level semantic relationships of all the input tokens. A subsequent linear layer serves as the generation head to model the probability distribution for the next token. During inference, we sample from this probability distribution to perform next-token generation. It is important to note that, since we have expanded the original vocabulary of the CLM, the generated tokens can belong to text, SMILES, or fragments. We initialize the weights of our model using the decoder layers from Llama-3.0-8B to improve training efficiency.

\subsection*{Training Procedure}
Our training process consists of two stages: alignment training and instruction tuning. 
During the alignment training stage, we aim to ensure that information $\textbf{U}_{uni}$ from various perspectives is well integrated and aligned. Thus, we freeze the AutoRegressive Decoder and train the remaining parameters $\Theta$, as shown in Fig.\ref{fig:model-archi}(a). Using the pretraining dataset $\mathcal{D}$, we encode the input molecules, requiring the AutoRegressive Decoder to generate the corresponding text. The alignment training loss is defined as follows:
\begin{equation}
    \mathcal{L}_{\text{align}}(\Theta) = \mathbb{E}_{\boldsymbol{x} \sim \mathcal{D}} \left[ - \sum_{i=1}^{n} \log p(x_i | \textbf{U}_{uni}, x_0, x_1, \ldots, x_{i-1}; \Theta) \right], 
\end{equation}
where $\boldsymbol{x} = \{x_0, x_1, \ldots, x_{n-1}\}$ represents the output sequence, with $x_i$ being the $i$-th token. $\mathcal{L}_{\text{align}}$ is the cross-entropy loss, minimizing the negative log-likelihood of the target sequence.

During the instruction tuning stage, we also involve the AutoRegressive Decoder in the training process, Fig.\ref{fig:model-archi}(b). Since the AutoRegressive Decoder has a large number of parameters, Low-Rank Adaptation (LoRA)~\cite{hu2021lora}, a parameter-efficient training method based on reparameterization~\cite{salimans2016weight,kingma2015variational}, is often used for fine-tuning. Specifically, training can be conceptualized as finding the change in parameters $\Delta\textbf{W}$. Hypothesizing $\Delta\textbf{W}$ is of low rank~\cite{li2018lowrank}, denoted as \(r\), it can be decomposed into two learnable low-rank matrices, $\Delta\textbf{W} = \textbf{AB}$. Thus the training objective can be transformed from finding $\Delta\textbf{W}$ to finding $\textbf{A}$ and $\textbf{B}$.


However, the performance of LoRA is inferior to full fine-tuning. Following the previous work~\cite{liudora}, $\textbf{W}$ can be decoupled as: $\textbf{W}=\textbf{m}\frac{\textbf{W}}{||\textbf{W}||_c}$, where $\textbf{m}$ is the amplitude vector, $||\cdot||_c$ represents the vector-wise norm of the matrix across each column, and $\frac{\textbf{W}}{||\textbf{W}||_c}$ is the direction matrix, with each column being a unit vector. We measure the changes in the amplitude and direction during training with the following formulas:
\begin{equation}
        \Delta \textbf{m}^t = \frac{\sum_{n=1}^k | \textbf{m}^{n,t} - \textbf{m}^{n,0} |}{k} , \quad \Delta \textbf{D}^t = \frac{\sum_{n=1}^{k} (1 - \textrm{cos}(\textbf{W}^{n,t}, \textbf{W}^{n,0}))}{k},
\end{equation}
where $n$ represents the $n$-th component, $\textrm{cos}$ represents the cosine calculation, $\Delta \textbf{m}^t$ denotes the amplitude change at the $t$-th step of training, and $\Delta \textbf{D}^t$ denotes the directional change. Previous work~\cite{liudora} has shown that, compared to full fine-tuning, LoRA lacks the ability to make fine adjustments to the direction matrix. Therefore, we update only the amplitude vector and apply low-rank decomposition solely to the direction matrix, allowing us to achieve better performance while maintaining efficient training.
\section*{Data availability}
Our MCMoD dataset is provided on \href{https://huggingface.co/datasets/GreatCaptainNemo/HME_dataset}{MCMoD}. The preprocessed data used for molecular comprehension tasks (pretraining, captioning, QA, docking) can refer to \href{https://huggingface.co/datasets/Sihangli/3D-MoIT}{3D-MoIT} and \href{https://github.com/aspuru-guzik-group/Tartarus}{Tartarus}.
All original databases used in this manuscript are publicly available and include \href{https://pubchem.ncbi.nlm.nih.gov/docs/downloads}{PubChem}, \href{https://nakatamaho.riken.jp/pubchemqc.riken.jp/}{PubChemQC}, \href{https://www.ebi.ac.uk/chebi/}{ChEBI}, \href{https://dtp.cancer.gov/}{DTP}, \href{https://zinc.docking.org/}{ZINC}.
To further facilitate reproducibility, we have integrated all datasets used in this manuscript into \href{https://huggingface.co/datasets/GreatCaptainNemo/HME_dataset}{HuggingFace} and provided the necessary data processing scripts on \href{https://github.com/Lyu6PosHao/HME}{https://github.com/Lyu6PosHao/HME}. 
\section*{Code availability}
All code used in data analysis and preparation of the manuscript, alongside a description of necessary steps for reproducing results, can be found in a GitHub repository accompanying this manuscript: \href{https://github.com/Lyu6PosHao/HME}{https://github.com/Lyu6PosHao/HME}.

\normalem
\bibliography{related_papers}

\begin{thebibliography}{10}
\urlstyle{rm}
\expandafter\ifx\csname url\endcsname\relax
  \def\url#1{\texttt{#1}}\fi
\expandafter\ifx\csname urlprefix\endcsname\relax\def\urlprefix{URL }\fi
\expandafter\ifx\csname doiprefix\endcsname\relax\def\doiprefix{DOI: }\fi
\providecommand{\bibinfo}[2]{#2}
\providecommand{\eprint}[2][]{\url{#2}}

\bibitem{restrepo2022chemical}
\bibinfo{author}{Restrepo, G.}
\newblock \bibinfo{journal}{\bibinfo{title}{Chemical space: limits, evolution and modelling of an object bigger than our universal library}}.
\newblock {\emph{\JournalTitle{Digital Discovery}}} \textbf{\bibinfo{volume}{1}}, \bibinfo{pages}{568--585} (\bibinfo{year}{2022}).

\bibitem{gromski2019explore}
\bibinfo{author}{Gromski, P.~S.}, \bibinfo{author}{Henson, A.~B.}, \bibinfo{author}{Granda, J.~M.} \& \bibinfo{author}{Cronin, L.}
\newblock \bibinfo{journal}{\bibinfo{title}{How to explore chemical space using algorithms and automation}}.
\newblock {\emph{\JournalTitle{Nature Reviews Chemistry}}} \textbf{\bibinfo{volume}{3}}, \bibinfo{pages}{119--128} (\bibinfo{year}{2019}).

\bibitem{wozniak2018linguistic}
\bibinfo{author}{Wo{\'z}niak, M.} \emph{et~al.}
\newblock \bibinfo{journal}{\bibinfo{title}{Linguistic measures of chemical diversity and the “keywords” of molecular collections}}.
\newblock {\emph{\JournalTitle{Scientific reports}}} \textbf{\bibinfo{volume}{8}}, \bibinfo{pages}{7598} (\bibinfo{year}{2018}).

\bibitem{zhou2024navigating}
\bibinfo{author}{Zhou, J.} \& \bibinfo{author}{Huang, M.}
\newblock \bibinfo{journal}{\bibinfo{title}{Navigating the landscape of enzyme design: from molecular simulations to machine learning}}.
\newblock {\emph{\JournalTitle{Chemical Society Reviews}}}  (\bibinfo{year}{2024}).

\bibitem{weininger1988smiles}
\bibinfo{author}{Weininger, D.}
\newblock \bibinfo{journal}{\bibinfo{title}{Smiles, a chemical language and information system. 1. introduction to methodology and encoding rules}}.
\newblock {\emph{\JournalTitle{Journal of chemical information and computer sciences}}} \textbf{\bibinfo{volume}{28}}, \bibinfo{pages}{31--36} (\bibinfo{year}{1988}).

\bibitem{krenn2020selfies}
\bibinfo{author}{Krenn, M.}, \bibinfo{author}{H{\"a}se, F.}, \bibinfo{author}{Nigam, A.}, \bibinfo{author}{Friederich, P.} \& \bibinfo{author}{Aspuru-Guzik, A.}
\newblock \bibinfo{journal}{\bibinfo{title}{Self-referencing embedded strings (selfies): A 100\% robust molecular string representation}}.
\newblock {\emph{\JournalTitle{Machine Learning: Science and Technology}}} \textbf{\bibinfo{volume}{1}}, \bibinfo{pages}{045024} (\bibinfo{year}{2020}).

\bibitem{molt5}
\bibinfo{author}{Edwards, C.} \emph{et~al.}
\newblock \bibinfo{title}{Translation between molecules and natural language}.
\newblock In \emph{\bibinfo{booktitle}{2022 Conference on Empirical Methods in Natural Language Processing, EMNLP 2022}} (\bibinfo{year}{2022}).

\bibitem{pei2023biot5}
\bibinfo{author}{Pei, Q.} \emph{et~al.}
\newblock \bibinfo{journal}{\bibinfo{title}{Biot5: Enriching cross-modal integration in biology with chemical knowledge and natural language associations}}.
\newblock {\emph{\JournalTitle{arXiv preprint arXiv:2310.07276}}}  (\bibinfo{year}{2023}).

\bibitem{fang2023molinst}
\bibinfo{author}{Fang, Y.} \emph{et~al.}
\newblock \bibinfo{journal}{\bibinfo{title}{Mol-instructions: A large-scale biomolecular instruction dataset for large language models}}.
\newblock {\emph{\JournalTitle{arXiv preprint arXiv:2306.08018}}}  (\bibinfo{year}{2023}).

\bibitem{elton2019deep}
\bibinfo{author}{Elton, D.~C.}, \bibinfo{author}{Boukouvalas, Z.}, \bibinfo{author}{Fuge, M.~D.} \& \bibinfo{author}{Chung, P.~W.}
\newblock \bibinfo{journal}{\bibinfo{title}{Deep learning for molecular design—a review of the state of the art}}.
\newblock {\emph{\JournalTitle{Molecular Systems Design \& Engineering}}} \textbf{\bibinfo{volume}{4}}, \bibinfo{pages}{828--849} (\bibinfo{year}{2019}).

\bibitem{brown2019guacamol}
\bibinfo{author}{Brown, N.}, \bibinfo{author}{Fiscato, M.}, \bibinfo{author}{Segler, M.~H.} \& \bibinfo{author}{Vaucher, A.~C.}
\newblock \bibinfo{journal}{\bibinfo{title}{Guacamol: benchmarking models for de novo molecular design}}.
\newblock {\emph{\JournalTitle{Journal of chemical information and modeling}}} \textbf{\bibinfo{volume}{59}}, \bibinfo{pages}{1096--1108} (\bibinfo{year}{2019}).

\bibitem{li2024empowering}
\bibinfo{author}{Li, J.} \emph{et~al.}
\newblock \bibinfo{journal}{\bibinfo{title}{Empowering molecule discovery for molecule-caption translation with large language models: A chatgpt perspective}}.
\newblock {\emph{\JournalTitle{IEEE Transactions on Knowledge and Data Engineering}}}  (\bibinfo{year}{2024}).

\bibitem{tran2024mol2lang}
\bibinfo{author}{Tran, D.}, \bibinfo{author}{Pham, N.~T.}, \bibinfo{author}{Nguyen, N.} \& \bibinfo{author}{Manavalan, B.}
\newblock \bibinfo{title}{Mol2lang-vlm: Vision-and text-guided generative pre-trained language models for advancing molecule captioning through multimodal fusion}.
\newblock In \emph{\bibinfo{booktitle}{Proceedings of the 1st Workshop on Language+ Molecules (L+ M 2024)}}, \bibinfo{pages}{97--102} (\bibinfo{year}{2024}).

\bibitem{qureshi2023ai}
\bibinfo{author}{Qureshi, R.} \emph{et~al.}
\newblock \bibinfo{journal}{\bibinfo{title}{Ai in drug discovery and its clinical relevance}}.
\newblock {\emph{\JournalTitle{Heliyon}}} \textbf{\bibinfo{volume}{9}} (\bibinfo{year}{2023}).

\bibitem{zongying2024taxdiff}
\bibinfo{author}{Zongying, L.} \emph{et~al.}
\newblock \bibinfo{journal}{\bibinfo{title}{Taxdiff: Taxonomic-guided diffusion model for protein sequence generation}}.
\newblock {\emph{\JournalTitle{arXiv preprint arXiv:2402.17156}}}  (\bibinfo{year}{2024}).

\bibitem{pei2024survey_biomolecule_llm}
\bibinfo{author}{Pei, Q.} \emph{et~al.}
\newblock \bibinfo{journal}{\bibinfo{title}{Leveraging biomolecule and natural language through multi-modal learning: A survey}}.
\newblock {\emph{\JournalTitle{arXiv preprint arXiv:2403.01528}}}  (\bibinfo{year}{2024}).

\bibitem{wang2022molr}
\bibinfo{author}{Wang, H.} \emph{et~al.}
\newblock \bibinfo{title}{Chemical-reaction-aware molecule representation learning}.
\newblock In \emph{\bibinfo{booktitle}{International Conference on Learning Representations}} (\bibinfo{year}{2022}).

\bibitem{zhou2023unimol}
\bibinfo{author}{Zhou, G.} \emph{et~al.}
\newblock \bibinfo{title}{Uni-mol: A universal 3d molecular representation learning framework}.
\newblock In \emph{\bibinfo{booktitle}{The Eleventh International Conference on Learning Representations}} (\bibinfo{year}{2023}).

\bibitem{zhang2024survey_scientific_llm}
\bibinfo{author}{Zhang, Q.} \emph{et~al.}
\newblock \bibinfo{journal}{\bibinfo{title}{Scientific large language models: A survey on biological \& chemical domains}}.
\newblock {\emph{\JournalTitle{arXiv preprint arXiv:2401.14656}}}  (\bibinfo{year}{2024}).

\bibitem{zhang2024survey_sci_llm_zhangyu}
\bibinfo{author}{Zhang, Y.} \emph{et~al.}
\newblock \bibinfo{journal}{\bibinfo{title}{A comprehensive survey of scientific large language models and their applications in scientific discovery}}.
\newblock {\emph{\JournalTitle{arXiv preprint arXiv:2406.10833}}}  (\bibinfo{year}{2024}).

\bibitem{moleculestm}
\bibinfo{author}{Liu, S.} \emph{et~al.}
\newblock \bibinfo{journal}{\bibinfo{title}{Multi-modal molecule structure--text model for text-based retrieval and editing}}.
\newblock {\emph{\JournalTitle{Nature Machine Intelligence}}} \textbf{\bibinfo{volume}{5}}, \bibinfo{pages}{1447--1457} (\bibinfo{year}{2023}).

\bibitem{cao2023instructmol}
\bibinfo{author}{Cao, H.}, \bibinfo{author}{Liu, Z.}, \bibinfo{author}{Lu, X.}, \bibinfo{author}{Yao, Y.} \& \bibinfo{author}{Li, Y.}
\newblock \bibinfo{journal}{\bibinfo{title}{Instructmol: Multi-modal integration for building a versatile and reliable molecular assistant in drug discovery}}.
\newblock {\emph{\JournalTitle{arXiv preprint arXiv:2311.16208}}}  (\bibinfo{year}{2023}).

\bibitem{3dmolm}
\bibinfo{author}{Li, S.} \emph{et~al.}
\newblock \bibinfo{journal}{\bibinfo{title}{Towards 3d molecule-text interpretation in language models}}.
\newblock {\emph{\JournalTitle{arXiv preprint arXiv:2401.13923}}}  (\bibinfo{year}{2024}).

\bibitem{skinnider2024invalidsmiles}
\bibinfo{author}{Skinnider, M.~A.}
\newblock \bibinfo{journal}{\bibinfo{title}{Invalid smiles are beneficial rather than detrimental to chemical language models}}.
\newblock {\emph{\JournalTitle{Nature Machine Intelligence}}} \textbf{\bibinfo{volume}{6}}, \bibinfo{pages}{437--448} (\bibinfo{year}{2024}).

\bibitem{chen2020oversmoothing}
\bibinfo{author}{Chen, D.} \emph{et~al.}
\newblock \bibinfo{title}{Measuring and relieving the over-smoothing problem for graph neural networks from the topological view}.
\newblock In \emph{\bibinfo{booktitle}{Proceedings of the AAAI conference on artificial intelligence}}, vol.~\bibinfo{volume}{34}, \bibinfo{pages}{3438--3445} (\bibinfo{year}{2020}).

\bibitem{xing2024overglobal}
\bibinfo{author}{Xing, Y.}, \bibinfo{author}{Wang, X.}, \bibinfo{author}{Li, Y.}, \bibinfo{author}{Huang, H.} \& \bibinfo{author}{Shi, C.}
\newblock \bibinfo{title}{Less is more: on the over-globalizing problem in graph transformers}.
\newblock In \emph{\bibinfo{booktitle}{Forty-first International Conference on Machine Learning}} (\bibinfo{year}{2024}).

\bibitem{crippen1988conformation}
\bibinfo{author}{Crippen, G.~M.}, \bibinfo{author}{Havel, T.~F.} \emph{et~al.}
\newblock \emph{\bibinfo{title}{Distance geometry and molecular conformation}}, vol.~\bibinfo{volume}{74} (\bibinfo{publisher}{Research Studies Press Taunton}, \bibinfo{year}{1988}).

\bibitem{liu2022graphmvp}
\bibinfo{author}{Liu, S.} \emph{et~al.}
\newblock \bibinfo{title}{Pre-training molecular graph representation with 3d geometry}.
\newblock In \emph{\bibinfo{booktitle}{International Conference on Learning Representations}} (\bibinfo{year}{2022}).

\bibitem{lyy-blending2d3d}
\bibinfo{author}{Yu, Q.} \emph{et~al.}
\newblock \bibinfo{title}{Multimodal molecular pretraining via modality blending}.
\newblock In \emph{\bibinfo{booktitle}{The Twelfth International Conference on Learning Representations}} (\bibinfo{year}{2024}).

\bibitem{lyy-unicorn2d3d}
\bibinfo{author}{Feng, S.} \emph{et~al.}
\newblock \bibinfo{journal}{\bibinfo{title}{Unicorn: A unified contrastive learning approach for multi-view molecular representation learning}}.
\newblock {\emph{\JournalTitle{arXiv preprint arXiv:2405.10343}}}  (\bibinfo{year}{2024}).

\bibitem{feng2024bioactivity}
\bibinfo{author}{Feng, B.} \emph{et~al.}
\newblock \bibinfo{journal}{\bibinfo{title}{A bioactivity foundation model using pairwise meta-learning}}.
\newblock {\emph{\JournalTitle{Nature Machine Intelligence}}} \textbf{\bibinfo{volume}{6}}, \bibinfo{pages}{962--974} (\bibinfo{year}{2024}).

\bibitem{khedkar2007pharmacophore}
\bibinfo{author}{Khedkar, S.~A.}, \bibinfo{author}{Malde, A.~K.}, \bibinfo{author}{Coutinho, E.~C.} \& \bibinfo{author}{Srivastava, S.}
\newblock \bibinfo{journal}{\bibinfo{title}{Pharmacophore modeling in drug discovery and development: an overview}}.
\newblock {\emph{\JournalTitle{Medicinal Chemistry}}} \textbf{\bibinfo{volume}{3}}, \bibinfo{pages}{187--197} (\bibinfo{year}{2007}).

\bibitem{van1982error3D}
\bibinfo{author}{Van~Eijck, B.}
\newblock \bibinfo{journal}{\bibinfo{title}{Influence of molecular vibrations on substitution coordinates}}.
\newblock {\emph{\JournalTitle{Journal of Molecular Spectroscopy}}} \textbf{\bibinfo{volume}{91}}, \bibinfo{pages}{348--362} (\bibinfo{year}{1982}).

\bibitem{tang2024cycle3d}
\bibinfo{author}{Tang, Z.} \emph{et~al.}
\newblock \bibinfo{journal}{\bibinfo{title}{Cycle3d: High-quality and consistent image-to-3d generation via generation-reconstruction cycle}}.
\newblock {\emph{\JournalTitle{arXiv preprint arXiv:2407.19548}}}  (\bibinfo{year}{2024}).

\bibitem{labute2005error3D}
\bibinfo{author}{Labute, P.}
\newblock \bibinfo{journal}{\bibinfo{title}{On the perception of molecules from 3d atomic coordinates}}.
\newblock {\emph{\JournalTitle{Journal of chemical information and modeling}}} \textbf{\bibinfo{volume}{45}}, \bibinfo{pages}{215--221} (\bibinfo{year}{2005}).

\bibitem{wei2022cot}
\bibinfo{author}{Wei, J.} \emph{et~al.}
\newblock \bibinfo{journal}{\bibinfo{title}{Chain-of-thought prompting elicits reasoning in large language models}}.
\newblock {\emph{\JournalTitle{Advances in neural information processing systems}}} \textbf{\bibinfo{volume}{35}}, \bibinfo{pages}{24824--24837} (\bibinfo{year}{2022}).

\bibitem{chu2023survey}
\bibinfo{author}{Chu, Z.} \emph{et~al.}
\newblock \bibinfo{journal}{\bibinfo{title}{A survey of chain of thought reasoning: Advances, frontiers and future}}.
\newblock {\emph{\JournalTitle{arXiv preprint arXiv:2309.15402}}}  (\bibinfo{year}{2023}).

\bibitem{mu2024embodiedgpt}
\bibinfo{author}{Mu, Y.} \emph{et~al.}
\newblock \bibinfo{journal}{\bibinfo{title}{Embodiedgpt: Vision-language pre-training via embodied chain of thought}}.
\newblock {\emph{\JournalTitle{Advances in Neural Information Processing Systems}}} \textbf{\bibinfo{volume}{36}} (\bibinfo{year}{2024}).

\bibitem{koppen2000cursedimension}
\bibinfo{author}{K{\"o}ppen, M.}
\newblock \bibinfo{title}{The curse of dimensionality}.
\newblock In \emph{\bibinfo{booktitle}{5th online world conference on soft computing in industrial applications (WSC5)}}, vol.~\bibinfo{volume}{1}, \bibinfo{pages}{4--8} (\bibinfo{year}{2000}).

\bibitem{honda2019smiles}
\bibinfo{author}{Honda, S.}, \bibinfo{author}{Shi, S.} \& \bibinfo{author}{Ueda, H.~R.}
\newblock \bibinfo{journal}{\bibinfo{title}{Smiles transformer: Pre-trained molecular fingerprint for low data drug discovery}}.
\newblock {\emph{\JournalTitle{arXiv preprint arXiv:1911.04738}}}  (\bibinfo{year}{2019}).

\bibitem{morgan1965morgan}
\bibinfo{author}{Morgan, H.~L.}
\newblock \bibinfo{journal}{\bibinfo{title}{The generation of a unique machine description for chemical structures-a technique developed at chemical abstracts service.}}
\newblock {\emph{\JournalTitle{Journal of chemical documentation}}} \textbf{\bibinfo{volume}{5}}, \bibinfo{pages}{107--113} (\bibinfo{year}{1965}).

\bibitem{bajusz2015tanimoto}
\bibinfo{author}{Bajusz, D.}, \bibinfo{author}{R{\'a}cz, A.} \& \bibinfo{author}{H{\'e}berger, K.}
\newblock \bibinfo{journal}{\bibinfo{title}{Why is tanimoto index an appropriate choice for fingerprint-based similarity calculations?}}
\newblock {\emph{\JournalTitle{Journal of cheminformatics}}} \textbf{\bibinfo{volume}{7}}, \bibinfo{pages}{1--13} (\bibinfo{year}{2015}).

\bibitem{li2024decoupled}
\bibinfo{author}{Li, H.} \emph{et~al.}
\newblock \bibinfo{journal}{\bibinfo{title}{Decoupled peak property learning for efficient and interpretable ecd spectra prediction}}.
\newblock {\emph{\JournalTitle{arXiv preprint arXiv:2401.03403}}}  (\bibinfo{year}{2024}).

\bibitem{wu2018moleculenet}
\bibinfo{author}{Wu, Z.} \emph{et~al.}
\newblock \bibinfo{journal}{\bibinfo{title}{Moleculenet: a benchmark for molecular machine learning}}.
\newblock {\emph{\JournalTitle{Chemical science}}} \textbf{\bibinfo{volume}{9}}, \bibinfo{pages}{513--530} (\bibinfo{year}{2018}).

\bibitem{luo2023airbiomedgpt}
\bibinfo{author}{Luo, Y.} \emph{et~al.}
\newblock \bibinfo{journal}{\bibinfo{title}{Biomedgpt: Open multimodal generative pre-trained transformer for biomedicine}}.
\newblock {\emph{\JournalTitle{arXiv preprint arXiv:2308.09442}}}  (\bibinfo{year}{2023}).

\bibitem{edwards2021text2molchebi20}
\bibinfo{author}{Edwards, C.}, \bibinfo{author}{Zhai, C.} \& \bibinfo{author}{Ji, H.}
\newblock \bibinfo{title}{Text2mol: Cross-modal molecule retrieval with natural language queries}.
\newblock In \emph{\bibinfo{booktitle}{Proceedings of the 2021 Conference on Empirical Methods in Natural Language Processing}}, \bibinfo{pages}{595--607} (\bibinfo{year}{2021}).

\bibitem{kang2018conditionalmolgen}
\bibinfo{author}{Kang, S.} \& \bibinfo{author}{Cho, K.}
\newblock \bibinfo{journal}{\bibinfo{title}{Conditional molecular design with deep generative models}}.
\newblock {\emph{\JournalTitle{Journal of chemical information and modeling}}} \textbf{\bibinfo{volume}{59}}, \bibinfo{pages}{43--52} (\bibinfo{year}{2018}).

\bibitem{kim2016pubchem}
\bibinfo{author}{Kim, S.} \emph{et~al.}
\newblock \bibinfo{journal}{\bibinfo{title}{Pubchem substance and compound databases}}.
\newblock {\emph{\JournalTitle{Nucleic acids research}}} \textbf{\bibinfo{volume}{44}}, \bibinfo{pages}{D1202--D1213} (\bibinfo{year}{2016}).

\bibitem{irwin2005zinc}
\bibinfo{author}{Irwin, J.~J.} \& \bibinfo{author}{Shoichet, B.~K.}
\newblock \bibinfo{journal}{\bibinfo{title}{Zinc- a free database of commercially available compounds for virtual screening}}.
\newblock {\emph{\JournalTitle{Journal of chemical information and modeling}}} \textbf{\bibinfo{volume}{45}}, \bibinfo{pages}{177--182} (\bibinfo{year}{2005}).

\bibitem{degtyarenko2007chebi}
\bibinfo{author}{Degtyarenko, K.} \emph{et~al.}
\newblock \bibinfo{journal}{\bibinfo{title}{Chebi: a database and ontology for chemical entities of biological interest}}.
\newblock {\emph{\JournalTitle{Nucleic acids research}}} \textbf{\bibinfo{volume}{36}}, \bibinfo{pages}{D344--D350} (\bibinfo{year}{2007}).

\bibitem{sorokina2021coconut}
\bibinfo{author}{Sorokina, M.}, \bibinfo{author}{Merseburger, P.}, \bibinfo{author}{Rajan, K.}, \bibinfo{author}{Yirik, M.~A.} \& \bibinfo{author}{Steinbeck, C.}
\newblock \bibinfo{journal}{\bibinfo{title}{Coconut online: collection of open natural products database}}.
\newblock {\emph{\JournalTitle{Journal of Cheminformatics}}} \textbf{\bibinfo{volume}{13}}, \bibinfo{pages}{2} (\bibinfo{year}{2021}).

\bibitem{monga2002dtp}
\bibinfo{author}{Monga, M.} \& \bibinfo{author}{Sausville, E.~A.}
\newblock \bibinfo{journal}{\bibinfo{title}{Developmental therapeutics program at the nci: molecular target and drug discovery process}}.
\newblock {\emph{\JournalTitle{Leukemia}}} \textbf{\bibinfo{volume}{16}}, \bibinfo{pages}{520--526} (\bibinfo{year}{2002}).

\bibitem{alhossary2015quickvina2}
\bibinfo{author}{Alhossary, A.}, \bibinfo{author}{Handoko, S.~D.}, \bibinfo{author}{Mu, Y.} \& \bibinfo{author}{Kwoh, C.-K.}
\newblock \bibinfo{journal}{\bibinfo{title}{Fast, accurate, and reliable molecular docking with quickvina 2}}.
\newblock {\emph{\JournalTitle{Bioinformatics}}} \textbf{\bibinfo{volume}{31}}, \bibinfo{pages}{2214--2216} (\bibinfo{year}{2015}).

\bibitem{koes2013smina}
\bibinfo{author}{Koes, D.~R.}, \bibinfo{author}{Baumgartner, M.~P.} \& \bibinfo{author}{Camacho, C.~J.}
\newblock \bibinfo{journal}{\bibinfo{title}{Lessons learned in empirical scoring with smina from the csar 2011 benchmarking exercise}}.
\newblock {\emph{\JournalTitle{Journal of chemical information and modeling}}} \textbf{\bibinfo{volume}{53}}, \bibinfo{pages}{1893--1904} (\bibinfo{year}{2013}).

\bibitem{2022zeng-kvplm}
\bibinfo{author}{Zeng, Z.}, \bibinfo{author}{Yao, Y.}, \bibinfo{author}{Liu, Z.} \& \bibinfo{author}{Sun, M.}
\newblock \bibinfo{journal}{\bibinfo{title}{A deep-learning system bridging molecule structure and biomedical text with comprehension comparable to human professionals}}.
\newblock {\emph{\JournalTitle{Nature communications}}} \textbf{\bibinfo{volume}{13}}, \bibinfo{pages}{862} (\bibinfo{year}{2022}).

\bibitem{momu}
\bibinfo{author}{Su, B.} \emph{et~al.}
\newblock \bibinfo{journal}{\bibinfo{title}{A molecular multimodal foundation model associating molecule graphs with natural language}}.
\newblock {\emph{\JournalTitle{arXiv preprint arXiv:2209.05481}}}  (\bibinfo{year}{2022}).

\bibitem{touvron2023llama2}
\bibinfo{author}{Touvron, H.} \emph{et~al.}
\newblock \bibinfo{journal}{\bibinfo{title}{Llama 2: Open foundation and fine-tuned chat models}}.
\newblock {\emph{\JournalTitle{arXiv preprint arXiv:2307.09288}}}  (\bibinfo{year}{2023}).

\bibitem{llama3modelcard}
\bibinfo{author}{Dubey, A.} \emph{et~al.}
\newblock \bibinfo{journal}{\bibinfo{title}{The llama 3 herd of models}}.
\newblock {\emph{\JournalTitle{arXiv preprint arXiv:2407.21783}}}  (\bibinfo{year}{2024}).

\bibitem{guo2023canllmchem}
\bibinfo{author}{Guo, T.} \emph{et~al.}
\newblock \bibinfo{journal}{\bibinfo{title}{What can large language models do in chemistry? a comprehensive benchmark on eight tasks}}.
\newblock {\emph{\JournalTitle{Advances in Neural Information Processing Systems}}} \textbf{\bibinfo{volume}{36}}, \bibinfo{pages}{59662--59688} (\bibinfo{year}{2023}).

\bibitem{ouyang2022instructgpt}
\bibinfo{author}{Ouyang, L.} \emph{et~al.}
\newblock \bibinfo{journal}{\bibinfo{title}{Training language models to follow instructions with human feedback}}.
\newblock {\emph{\JournalTitle{Advances in neural information processing systems}}} \textbf{\bibinfo{volume}{35}}, \bibinfo{pages}{27730--27744} (\bibinfo{year}{2022}).

\bibitem{sherstinsky2020rnn}
\bibinfo{author}{Sherstinsky, A.}
\newblock \bibinfo{journal}{\bibinfo{title}{Fundamentals of recurrent neural network (rnn) and long short-term memory (lstm) network}}.
\newblock {\emph{\JournalTitle{Physica D: Nonlinear Phenomena}}} \textbf{\bibinfo{volume}{404}}, \bibinfo{pages}{132306} (\bibinfo{year}{2020}).

\bibitem{vaswani2017transformer}
\bibinfo{author}{Vaswani, A.}
\newblock \bibinfo{journal}{\bibinfo{title}{Attention is all you need}}.
\newblock {\emph{\JournalTitle{Advances in Neural Information Processing Systems}}}  (\bibinfo{year}{2017}).

\bibitem{liu2023molxpt}
\bibinfo{author}{Liu, Z.} \emph{et~al.}
\newblock \bibinfo{title}{Molxpt: Wrapping molecules with text for generative pre-training}.
\newblock In \emph{\bibinfo{booktitle}{Proceedings of the 61st Annual Meeting of the Association for Computational Linguistics (Volume 2: Short Papers)}}, \bibinfo{pages}{1606--1616} (\bibinfo{year}{2023}).

\bibitem{yang2024digfrag}
\bibinfo{author}{Yang, R.}, \bibinfo{author}{Zhou, H.}, \bibinfo{author}{Wang, F.} \& \bibinfo{author}{Yang, G.}
\newblock \bibinfo{journal}{\bibinfo{title}{Digfrag as a digital fragmentation method used for artificial intelligence-based drug design}}.
\newblock {\emph{\JournalTitle{Communications Chemistry}}} \textbf{\bibinfo{volume}{7}}, \bibinfo{pages}{258} (\bibinfo{year}{2024}).

\bibitem{preuer2018fcd}
\bibinfo{author}{Preuer, K.}, \bibinfo{author}{Renz, P.}, \bibinfo{author}{Unterthiner, T.}, \bibinfo{author}{Hochreiter, S.} \& \bibinfo{author}{Klambauer, G.}
\newblock \bibinfo{journal}{\bibinfo{title}{Fr{\'e}chet chemnet distance: a metric for generative models for molecules in drug discovery}}.
\newblock {\emph{\JournalTitle{Journal of chemical information and modeling}}} \textbf{\bibinfo{volume}{58}}, \bibinfo{pages}{1736--1741} (\bibinfo{year}{2018}).

\bibitem{hu2021lora}
\bibinfo{author}{Hu, E.~J.} \emph{et~al.}
\newblock \bibinfo{journal}{\bibinfo{title}{Lora: Low-rank adaptation of large language models}}.
\newblock {\emph{\JournalTitle{arXiv preprint arXiv:2106.09685}}}  (\bibinfo{year}{2021}).

\bibitem{salimans2016weight}
\bibinfo{author}{Salimans, T.} \& \bibinfo{author}{Kingma, D.~P.}
\newblock \bibinfo{journal}{\bibinfo{title}{Weight normalization: A simple reparameterization to accelerate training of deep neural networks}}.
\newblock {\emph{\JournalTitle{Advances in neural information processing systems}}} \textbf{\bibinfo{volume}{29}} (\bibinfo{year}{2016}).

\bibitem{kingma2015variational}
\bibinfo{author}{Kingma, D.~P.}, \bibinfo{author}{Salimans, T.} \& \bibinfo{author}{Welling, M.}
\newblock \bibinfo{journal}{\bibinfo{title}{Variational dropout and the local reparameterization trick}}.
\newblock {\emph{\JournalTitle{Advances in neural information processing systems}}} \textbf{\bibinfo{volume}{28}} (\bibinfo{year}{2015}).

\bibitem{li2018lowrank}
\bibinfo{author}{Li, Y.}, \bibinfo{author}{Ma, T.} \& \bibinfo{author}{Zhang, H.}
\newblock \bibinfo{title}{Algorithmic regularization in over-parameterized matrix sensing and neural networks with quadratic activations}.
\newblock In \emph{\bibinfo{booktitle}{Conference On Learning Theory}}, \bibinfo{pages}{2--47} (\bibinfo{organization}{PMLR}, \bibinfo{year}{2018}).

\bibitem{liudora}
\bibinfo{author}{Liu, S.-y.} \emph{et~al.}
\newblock \bibinfo{title}{Dora: Weight-decomposed low-rank adaptation}.
\newblock In \emph{\bibinfo{booktitle}{Forty-first International Conference on Machine Learning}} (\bibinfo{year}{2024}).

\bibitem{orgmoldesign}
\bibinfo{author}{Chen, Z.} \emph{et~al.}
\newblock \bibinfo{journal}{\bibinfo{title}{Multi-granularity score-based generative framework enables efficient inverse design of complex organics}}.
\newblock {\emph{\JournalTitle{arXiv preprint arXiv:2409.07912}}}  (\bibinfo{year}{2024}).

\bibitem{degen2008brics}
\bibinfo{author}{Degen, J.}, \bibinfo{author}{Wegscheid-Gerlach, C.}, \bibinfo{author}{Zaliani, A.} \& \bibinfo{author}{Rarey, M.}
\newblock \bibinfo{journal}{\bibinfo{title}{On the art of compiling and using'drug-like'chemical fragment spaces}}.
\newblock {\emph{\JournalTitle{ChemMedChem}}} \textbf{\bibinfo{volume}{3}}, \bibinfo{pages}{1503} (\bibinfo{year}{2008}).

\bibitem{lewell1998recap}
\bibinfo{author}{Lewell, X.~Q.}, \bibinfo{author}{Judd, D.~B.}, \bibinfo{author}{Watson, S.~P.} \& \bibinfo{author}{Hann, M.~M.}
\newblock \bibinfo{journal}{\bibinfo{title}{Recap retrosynthetic combinatorial analysis procedure: a powerful new technique for identifying privileged molecular fragments with useful applications in combinatorial chemistry}}.
\newblock {\emph{\JournalTitle{Journal of chemical information and computer sciences}}} \textbf{\bibinfo{volume}{38}}, \bibinfo{pages}{511--522} (\bibinfo{year}{1998}).

\bibitem{kong2022psvae}
\bibinfo{author}{Kong, X.}, \bibinfo{author}{Huang, W.}, \bibinfo{author}{Tan, Z.} \& \bibinfo{author}{Liu, Y.}
\newblock \bibinfo{journal}{\bibinfo{title}{Molecule generation by principal subgraph mining and assembling}}.
\newblock {\emph{\JournalTitle{Advances in Neural Information Processing Systems}}} \textbf{\bibinfo{volume}{35}}, \bibinfo{pages}{2550--2563} (\bibinfo{year}{2022}).

\bibitem{micam}
\bibinfo{author}{Geng, Z.} \emph{et~al.}
\newblock \bibinfo{title}{De novo molecular generation via connection-aware motif mining}.
\newblock In \emph{\bibinfo{booktitle}{The Eleventh International Conference on Learning Representations}} (\bibinfo{year}{2023}).

\bibitem{wu2024tsmiles}
\bibinfo{author}{Wu, J.-N.} \emph{et~al.}
\newblock \bibinfo{journal}{\bibinfo{title}{t-smiles: a fragment-based molecular representation framework for de novo ligand design}}.
\newblock {\emph{\JournalTitle{Nature Communications}}} \textbf{\bibinfo{volume}{15}}, \bibinfo{pages}{4993} (\bibinfo{year}{2024}).

\bibitem{ramesh2021dalle}
\bibinfo{author}{Ramesh, A.} \emph{et~al.}
\newblock \bibinfo{title}{Zero-shot text-to-image generation}.
\newblock In \emph{\bibinfo{booktitle}{International conference on machine learning}}, \bibinfo{pages}{8821--8831} (\bibinfo{organization}{Pmlr}, \bibinfo{year}{2021}).

\bibitem{van2017vqvae}
\bibinfo{author}{Van Den~Oord, A.}, \bibinfo{author}{Vinyals, O.} \emph{et~al.}
\newblock \bibinfo{journal}{\bibinfo{title}{Neural discrete representation learning}}.
\newblock {\emph{\JournalTitle{Advances in neural information processing systems}}} \textbf{\bibinfo{volume}{30}} (\bibinfo{year}{2017}).

\end{thebibliography}

\newpage
\section*{Supplementary Materials}
\renewcommand{\figurename}{Supplementary Figure}
\renewcommand{\tablename}{Supplementary Table}
\setcounter{figure}{0} 
\setcounter{table}{0} 
\subsection*{Supplementary Section~1: Evaluation of Molecular Encoding Bias}
We used 2,000 molecules from the test dataset of the captioning task to measure encoding bias. Molecular 1D, 2D, and 3D features were extracted using SciBERT, GIN, and UniMol, respectively. Global molecular features were obtained by applying mean pooling. Morgan fingerprints were extracted using RDKit, with 2,048 bits and a radius of 2. Additionally, we applied mean pooling to the first hidden layer features of HME to derive our molecular features. The statistics of molecular features are summarized in Supplementary~Table.~\ref{tab:encoding-statistic}:

\begin{table}[h]
\centering
\footnotesize
\caption{\textbf{Details of different molecular encodings.} There are significant differences in the dimensions and numerical types of different encodings.}
\label{tab:encoding-statistic}
\begin{tabular}{c|c|c|c}
\toprule
 &\textbf{Encoding} & \textbf{Dimension} & \textbf{Type} \\ \midrule
\textbf{1D}&SciBERT   & 768   & Float Vector        \\ \midrule
\textbf{2D}&GIN   & 300   & Float Vector     \\ \midrule 
\textbf{3D}&UniMol  & 512  & Float Vector      \\\midrule
\textbf{Ours}&HME  & 4096  & Float Vector   \\ \midrule
\textbf{Refer}&Morgan  & 2048  & Int Vector  \\ \bottomrule
\end{tabular}
\end{table}

Then we calculated the similarity matrix and analyzed the correlation coefficients between different matrices. Specifically, For 1D, 2D, 3D, and our features, we employed cosine similarity to measure the similarity between molecular pairs $[\mathcal{M}_i$, $\mathcal{M}_j]$ from the molecule dataset $\mathcal{M}$:
\begin{equation}
\textrm{sim}_{cos}(\mathcal{M}_i, \mathcal{M}_j) = \frac{\textbf{v}_i \cdot \textbf{v}_j}{||\textbf{v}_i|| \cdot ||\textbf{v}_j||},
\end{equation}
where $\textbf{v}_i$ and $\textbf{v}_j$ represent the feature vectors of molecules $\mathcal{M}_i$ and $\mathcal{M}_j$, respectively. For Morgan fingerprints, which are discrete integer values, we employed Tanimoto similarity instead:
\begin{equation}
\textrm{sim}_{tani}(\mathcal{M}_i, \mathcal{M}_j) = \frac{|\textbf{f}_i \cap \textbf{f}_j|}{|\textbf{f}_i \cup \textbf{f}_j|},
\end{equation}
where $\textbf{f}_i$ and $\textbf{f}_j$ denote the fingerprint bit vectors of molecules $\mathcal{M}_i$ and $\mathcal{M}_j$.
For each encoding method $k$, we construct a similarity matrix $\textbf{S}^k$ for the molecule set $\mathcal{M}$:
\begin{equation}
\textbf{S}^k = [s^k_{ij}]_{n \times n}, \quad s^k_{ij} = \textrm{sim}(\mathcal{M}_i, \mathcal{M}_j),
\end{equation}
where $n$ is the total number of molecules in $\mathcal{M}$, and $s^k_{ij}$ represents the similarity between molecules $\mathcal{M}_i$ and $\mathcal{M}_j$ under encoding method $k$.

Finally, we analyzed the Pearson correlation coefficients between pairs of similarity matrices ($\textbf{S}^p$ and $\textbf{S}^q$) obtained from different encoding methods:
\begin{equation}
\rho(\textbf{S}^p, \textbf{S}^q) = \frac{\sum_{i,j} (s^p_{ij} - \bar{s^p})(s^q_{ij} - \bar{s^q})}{\sqrt{\sum_{i,j} (s^p_{ij} - \bar{s^p})^2} \sqrt{\sum_{i,j} (s^q_{ij} - \bar{s^q})^2}},
\end{equation}
where $\bar{s^p}$ and $\bar{s^q}$ denote the mean values of all elements in matrices $\textbf{S}^p$ and $\textbf{S}^q$, respectively. This correlation analysis helps us understand the consistency and differences among various molecular encoding strategies, as shown in Supplementary~Fig.~\ref{fig:heatmap}. 

\begin{figure}[h]
  \centering
  \includegraphics[width=0.4\linewidth]{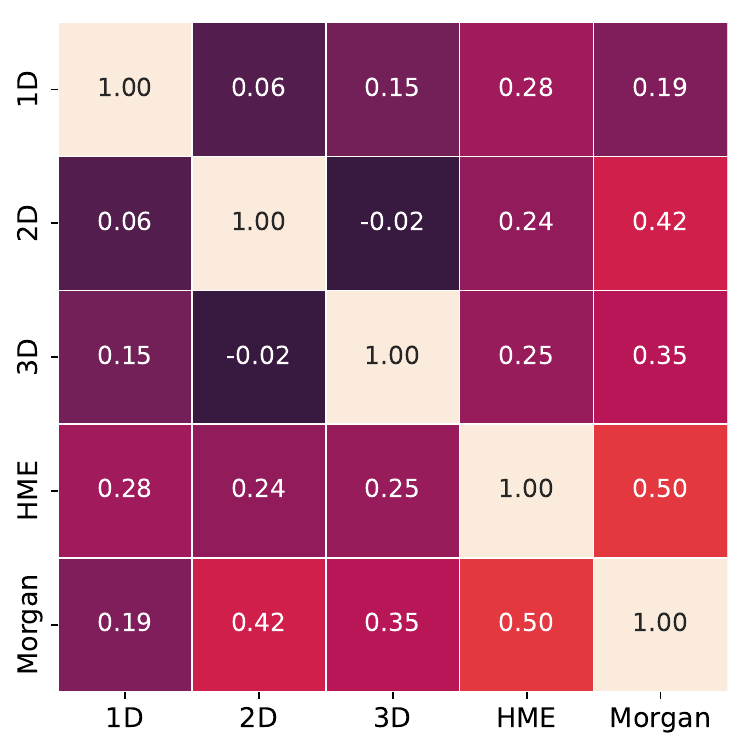}
  \caption{\textbf{Pearson Correlation between different molecular encodings.} Morgan: Morgan Fingerprints. Given a molecule set, we calculate the similarity matrix between molecular pairs using 5 different encodings~(1D,2D,3D, our HME, and Morgan). Then we calculate the Pearson correlation among these 5 similarity matrices.}
  \label{fig:heatmap}
\end{figure}

\subsection*{Supplementary Section~2: Dataset Construction and Composition}


In the molecular comprehension field, the 3D-MoIT dataset was adopted, which was constructed from PubChem and PubChemQC databases. 301K pairs were used for pretraining to align molecular and text modalities. For the captioning task, 12K molecule-text pairs were utilized as the training dataset and 2K as the test dataset. For molecular general QA, five QA pairs per molecule were generated using GPT-3.5 based on PubChem descriptions, focusing on physical-chemical properties, origins, and applications, resulting in 60K QA pairs for training and 10K for testing. The computational QA pairs were constructed in two parts: (1) four common properties from PubChem (molecular weight, LogP, TPSA, and complexity), yielding 46.7K QA pairs for training and 7.8K for testing; (2) four quantum chemical properties from PubChemQC (HOMO, LUMO, HOMO-LUMO Gap, and SCF Energy), resulting in 2.5M QA pairs for training and 312K for testing. More details about the dataset construction can be found in 3D-MoIT~\cite{3dmolm}. For the molecular docking score prediction task, a subset of the Tartarus dataset was adopted, which contains 100K molecules selected from the DTP database and their docking scores with protein PDB 4LDE. The dataset was split into training and test sets with a ratio of 9:1.

Molecular generation has become a cornerstone of modern drug discovery and material design, and the development of datasets that facilitate condition-driven molecule design plays a pivotal role in advancing this field. In this context, we introduce the MCMoD, a comprehensive resource tailored to support molecular design based on diverse conditioning factors, as shown in Supplementary~Table~.\ref{tab:MCMoD}. These conditions encompass textual descriptions, molecular properties (e.g., LogP, QED, SAS, and docking scores), and molecular fragments, providing a versatile framework for addressing a wide range of molecular design tasks. The target molecules in MCMoD span synthetic compounds, natural product-like molecules, and protein ligands, highlighting the dataset’s broad applicability and relevance to both synthetic chemistry and biochemistry.

The MCMoD dataset integrates data from multiple high-quality molecular sources, each contributing unique strengths and diversity to the overall corpus. Specifically, we obtained 250K synthetic molecules from the ZINC-250K dataset. For natural product-like molecules, we incorporated 695K molecules from the COCONUT dataset, which is one of the most extensive repositories of natural compounds, enriched with property annotations. Additionally, we leveraged a curated set of 100K drug-like molecules from the Tartarus dataset, which includes docking scores obtained using QuickVina-2 for docking pose sampling and SMINA for redocking. To enable molecule design based on textual descriptions, we combined data from ChEBI-20 and PubChem, resulting in a combined set of 332K molecules paired with detailed textual descriptions. For all the obtained molecules, we used RDKit to calculate molecular properties including LogP, QED, and SAS values, which are crucial for practical applications such as drug design. We quantized LogP and SAS values with a step size of 1.0, and QED with a step size of 0.1, to help the CLMs better learn the chemical landscape implied by the magnitudes of these property values. We used our fragment generator to fragment each molecule. When fragment sequences serve as Chains of Thought (CoT), we sorted them lexicographically to mitigate the adverse effects of sequence order. When fragments serve as conditions, we deduplicated the fragments for each molecule and excluded fragments that are composed solely of carbon atoms and lack conjugated structures. Finally, we randomly selected 1 to 3 fragments to serve as the condition.

MCMoD is designed to support a diverse set of molecular design tasks, with a particular emphasis on the role of molecular fragments. Key tasks include description-based molecular generation, where models synthesize molecules conditioned on textual inputs; multi-objective molecular reverse design, where molecules are optimized to meet multiple property criteria; and affinity-based ligand generation, which involves designing ligands with high binding affinities for specific protein targets. Molecular fragments play a dual role across these tasks: as CoT, fragments are sequentially generated by models to refine the molecular design process, whereas, in prompt-based settings, fragments serve as explicit conditioning inputs, guiding the generative process toward specific structural motifs or functionalities. We formulated multiple conditions in natural language, which enhances the model's ability to generalize across diverse scenarios.

In summary, the MCMoD dataset represents a significant step forward in conditional molecular generation by integrating diverse molecular sources and advanced conditioning paradigms, offering researchers a rich and flexible resource to tackle challenges in synthetic chemistry, natural product discovery, and ligand design.


\begin{table}[t]
\centering
\footnotesize
\caption{\textbf{Statistics of the MCMoD dataset.} Desc: Description. Prop: Property~(LogP, QED, SAS, additional docking score for ligands in DTP). Frag: Fragment. *: Molecules that overlapped with the ChEBI test dataset were filtered out to prevent data leakage. CoT: The fragment sequences are generated by models as Chain of Thought. Prompt: The fragments are provided to models as conditions.}
\label{tab:MCMoD}
\begin{tabular}{c|c|c|c|c}
\toprule
\textbf{Source} & \textbf{Size} & \textbf{Composition} & \textbf{Task} & \textbf{Role of Frag} \\ \midrule
PubChem   & 299K*   & Desc, Prop, Frag, SMILES   & Description-based Molecular Generation   & CoT     \\ \midrule
ChEBI   & 33K   & Desc, Prop, Frag, SMILES  & Description-based Molecular Generation  & CoT    \\ \midrule 
ZINC  & 250K  & Prop, Frag, SMILES  & Multi-objective Molecular Reverse Design  & Prompt    \\\midrule
COCONUT  & 695K  & Prop, Frag, SMILES  & Multi-objective Molecular Reverse Design  & Prompt   \\ \midrule
DTP  & 100K  & Prop, Frag, SMILES  & Affinity-Based Ligand Generation   & CoT  \\ \bottomrule
\end{tabular}
\end{table}

\subsection*{Supplementary Section~3: Prompt Format of HME}

\begin{table}[t]
\centering
\footnotesize
\caption{\textbf{The prompt format of our HME on each task.} The content within \{\} represents placeholders that vary depending on each specific sample. \{Features\} is filled by concatenating embeddings, while others are directly filled with strings.}
\label{tab:prompt-format}
\begin{tabular}{c|p{10cm}}
\toprule
\textbf{Task} & \textbf{Prompt}  \\ \midrule
Molecular Captioning   &  Please describe the molecule: Molecular geometric features are: \{Features\}. Molecular SMILES is \{SMILES\}. Molecular fragments are \{Fragments\}.      \\ \midrule
Molecular General QA   & \{Question\}. Molecular geometric features are: \{Features\}. Molecular SMILES is \{SMILES\}. Molecular fragments are \{Fragments\}.        \\ \midrule 
Molecular Property QA  & \{Question\}. Molecular geometric features are: \{Features\}. Molecular SMILES is \{SMILES\}. Molecular fragments are \{Fragments\}.     \\ \midrule
Protein-Ligand Affinity Prediction  & I am interested in the docking score of the molecule to Protein 4lde, could you tell me what it is? If uncertain, provide an estimate. Respond with the numerical value only. Molecular geometric features are: \{Features\}. Molecular SMILES is \{SMILES\}. Molecular fragments are \{Fragments\}.     \\ \midrule
Description-based Molecular Generation  & Please give me molecular fragments based on the description. And then give me the molecular SMILES based on both the fragments and the description. The description is: \{Description\}   \\ \midrule
Multi-objective Molecular Reverse Design & There are some conditions, including logp (the hydrophobicity and solubility balance), qed (the drug-likeness), sas (the synthetic accessibility score), and the fragments (include specific fragments). Now please design a molecule under the given constraints: The molecule should have these fragments \{Fragments\}. The molecule should have a \{Property Type\} value of \{Property Value\}. \\ \midrule
Affinity-Based Ligand Generation & Please give me molecular fragments based on the description. And then give me the molecular SMILES based on both the fragments and the description. The description is: The docking score of the molecule to Protein 4lde is \{Value\}.    \\ \bottomrule
\end{tabular}
\end{table}
To enhance the reproducibility, we listed the prompts used for each task in Supplementary~Table.~\ref{tab:prompt-format}. The content within {} is populated according to each sample. Notably, the Features are populated at the embedding level rather than the string level. In the Molecular General QA task, {Question} includes queries such as ``What are some of the physical properties of 2-Phenylethylamine?” and ``What are the roles of glycodihydroceramides in cellular processes?”. In the Molecule Property QA task, {Question} encompasses queries such as ``I need to know the LogP of this molecule, could you please provide it? If uncertain, provide an estimate. Respond with the numerical value only.” and ``Could you give me the HOMO-LUMO Gap value of this molecule? If uncertain, provide an estimate. Respond with the numerical value only.”

\subsection*{Supplementary Section~4: Detailed Results of the Protein-Ligand Affinity Prediction Task}

\begin{figure}[t]
  \centering
  \includegraphics[width=0.7\textwidth]{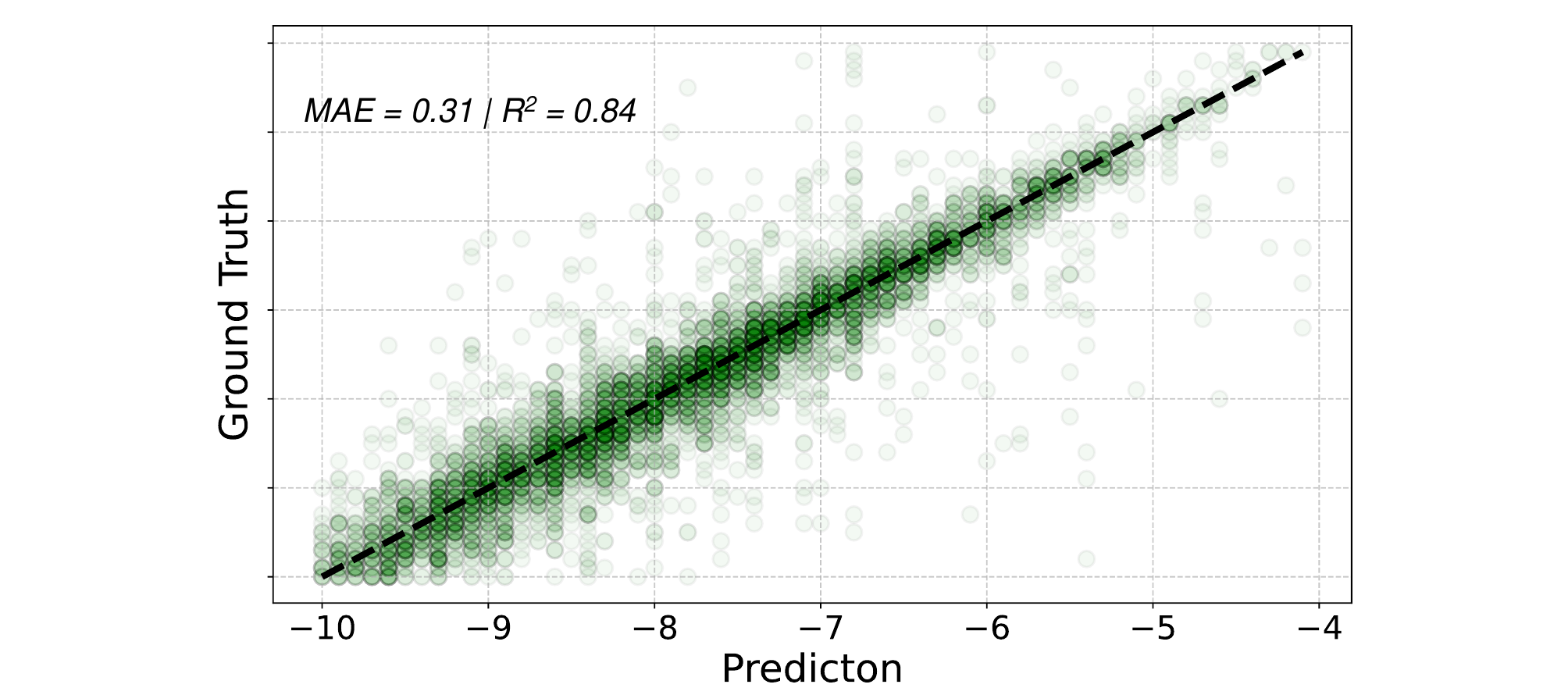}
  \caption{\textbf{The distribution of groud-truth and HME's prediction for docking score.} The dotted line is the diagonal line, indicating that the ground truth value and the predicted value are equal. MAE denotes the mean absolute error; R$^2$ denotes the correlation coefficient.}
  \label{fig:predict_docking}
\end{figure}

This task is formulated as follows: the model receives molecular representations as input and predicts the docking score of the molecule with a given protein. We extract the numerical values of the properties using regular expressions and then compare them with the true values.
As shown in Supplementary~Fig.~\ref{fig:predict_docking}, most data points are distributed around the diagonal line (indicated by the dashed line), representing a perfect correlation between predictions and true values. The MAE is 0.31, and R$^2$ is 0.84, indicating that the prediction model approximates true binding affinities well. This demonstrates that our HME has a certain understanding of protein-ligand interactions, facilitating drug screening.


\subsection*{Supplementary Section~5: Detailed Results of the Molecular Property QA Task}

\begin{table}[t]
\centering
\footnotesize
\caption{\textbf{Performance for the molecular property QA task.} We propose the experimental results on our HME and the corresponding baselines including CLMs, LLMs, and Uni-Mol. MAE: Mean absolute error. Valid: The answer of the LMs contains property values and the values are within a reasonable range. R$^2$: The coefficient of determination measuring how well the predictions fit the ground truth, ranging from 0 to 1.}
\label{tab:property-qa}
 \resizebox{\textwidth}{!}{
\begin{tabular}{llcccccccc}
\toprule
\textbf{Metrics} &\textbf{ Model} &\textbf{Weight (g/mol)}&\textbf{LogP}&\textbf{TPSA (\text{\AA}$^2$)}&\textbf{Complexity}&\textbf{HOMO (eV)}&\textbf{LUMO (eV)}&{\textbf{H-L Gap (eV)}}&{\textbf{SCF ($10^4$eV)}}\\
\midrule
 & {Uni-Mol} & {20.35} &{0.59}&{13.48}&{57.24}&{0.32}&{0.35}&{0.21}&{0.45}\\
\multirow{3}{*}{MAE$(\downarrow)$}
&Llama2-7B        &  22.10& 1.45 & 15.87 &  69.74 & 1.24   & 1.04& 0.88 & 0.70\\
&2D-MoLM        & 21.48 & 0.88  &  13.52& 55.74  & 0.92  & 0.80& 0.67  & 0.71\\
&3D-MoLM        & 14.79  & 0.66& 9.71 & {44.85}  & {0.26}  &  {0.25} & {0.28} & {0.35} \\
&{HME}      & \textbf{8.35}  & \textbf{0.51}  &  \textbf{5.61}  & \textbf{27.00}  & \textbf{0.19}  & \textbf{0.19}  &\textbf{0.19}& \textbf{0.01} \\
\midrule
\multirow{4}{*}{Valid$(\uparrow)$}
&Llama2-7B        & 96\% & 95\% &92\% &  93\% & 96\%  & 95\% & 92\% & 99\% \\
&2D-MoLM        & 94\% & 96\% &  92\% & 94\%  & 98\%  & 96\% & 93\% & 99\%\\
&3D-MoLM        & 95\% & 97\% & 93\% & 94\% & 97\% &  94\% & 94\% & 99\%\\
 & HME      & \textbf{99.90\%} &\textbf{100\%} &  \textbf{99.90\%}& \textbf{99.90\%}  &\textbf{ 100\%}  & \textbf{100\%} & \textbf{100\%} & \textbf{99.33\%} \\
\midrule
\multirow{1}{*}{R$^2(\uparrow)$}
 & 3D-MoLM & 0.9796 & 0.9000 & 0.8583 & 0.9474 & 0.0588 & 0.5893 & 0.8366 & 0.9635 \\
 & HME      & \textbf{0.9953} & \textbf{0.9692} &  \textbf{0.9898} & \textbf{0.9871}  & \textbf{0.6255} & \textbf{0.8138} & \textbf{0.9410} & \textbf{0.9994}  \\
\bottomrule
\addlinespace[0.1cm]
\end{tabular}}
\end{table}

\begin{table}[t]
\centering
\footnotesize
\caption{\textbf{Value ranges for molecular properties.} We define the reasonable ranges for different molecular properties to determine the validity of model predictions. Sup and Inf represent the lower and upper bounds respectively. Numbers marked with * indicate closed intervals (inclusive bounds).}
\label{tab:property-range}
\setlength{\tabcolsep}{3pt}
\begin{tabular}{lcccccccc}
\toprule
 &\textbf{Weight (g/mol)}&\textbf{LogP}&\textbf{TPSA (\text{\AA}$^2$)}&\textbf{Complexity}&\textbf{HOMO (eV)}&\textbf{LUMO (eV)}&{\textbf{H-L Gap (eV)}}&{\textbf{SCF ($10^4$eV)}}\\
\midrule
\textbf{Sup} & 0 & -30 & 0* & 0* & -20 & -20 & -20 & -50 \\
\textbf{Inf} & 4000 & 50 & 2000 & 10000* & 20 & 20 & 20 & 0 \\
\bottomrule
\addlinespace[0.1cm]
\end{tabular}
\end{table}

In Supplementary~Table.~\ref{tab:property-qa}, we present the quantitative performance of our HME on the Property QA task. In addition to conventional metrics such as MAE and R$^2$, we introduce another metric, the Valid Ratio. It is necessary because LMs generate property predictions in natural language, which may deviate from expectations, such as failing to include the anticipated property value or producing values outside a reasonable range (e.g., a negative molecular weight). Specifically, a valid answer $\mathcal{A}$ is defined as follows: (1) $\mathcal{A}$ contains a property value $\mathcal{V}$. (2) $\mathcal{V}$ lies within a reasonable range, as shown in Supplementary~Table.~\ref{tab:property-range}.

As shown in Supplementary~Table.~\ref{tab:property-qa}, HME demonstrates superior performance in answer validity compared to baseline models. It achieves a 100\% validity rate for LogP, HOMO, LUMO, and H-L Gap, with nearly 100\% validity for other properties. This indicates that HME successfully learns reasonable ranges of property values, reflecting its solid grasp of chemical knowledge. In terms of the Mean Absolute Error~(MAE), HME achieves the lowest average error; however, MAE only reflects the overall error level and does not capture the correlation between predicted and true values. From the perspective of the R$^2$ metric, 3D-MoLM shows an R$^2$ value of merely 0.0588 for the HOMO property, suggesting that its predictions are almost uncorrelated with the ground truth and akin to random guesses. In contrast, HME achieves R$^2$ values exceeding 0.9 for six properties and above 0.6 for all eight, indicating its ability to accurately capture trends in molecular properties and establish strong linear correlations. These results highlight the critical support that HME provides for the exploration and analysis of chemical properties.

\subsection*{Supplementary Section~6: Ablation Study for HME}
\textbf{Different Sizes of HME.}
We conduct experiments to evaluate the performance of HME variants with different model sizes on two fundamental tasks: molecular captioning and general QA. As shown in Supplementary~Table.~\ref{tab:different-HME}, we compare three versions: HME-Small, HME-Medium, and HME of the standard version. For the captioning task, HME-Small achieves BLEU-4 of 23.94 and ROUGE-L of 34.31, while HME-Medium improves these metrics to 24.38 and 35.6 respectively. HME further boosts the performance to BLEU-4 of 27.79 and ROUGE-L of 37.27. This trend is even more pronounced in the general QA task, where HME-Small obtains BLEU-4 of 35.89 and METEOR of 53.69, HME-Medium achieves 42.42 and 57.75, and HME significantly outperforms both with BLEU-4 of 46.41 and METEOR of 60.15. The performance gap between models is substantial, suggesting that a larger decoder enables HME to better capture the complex relationships between molecules and textual descriptions.

\begin{figure}[t]
  \centering
  \includegraphics[width=\textwidth]{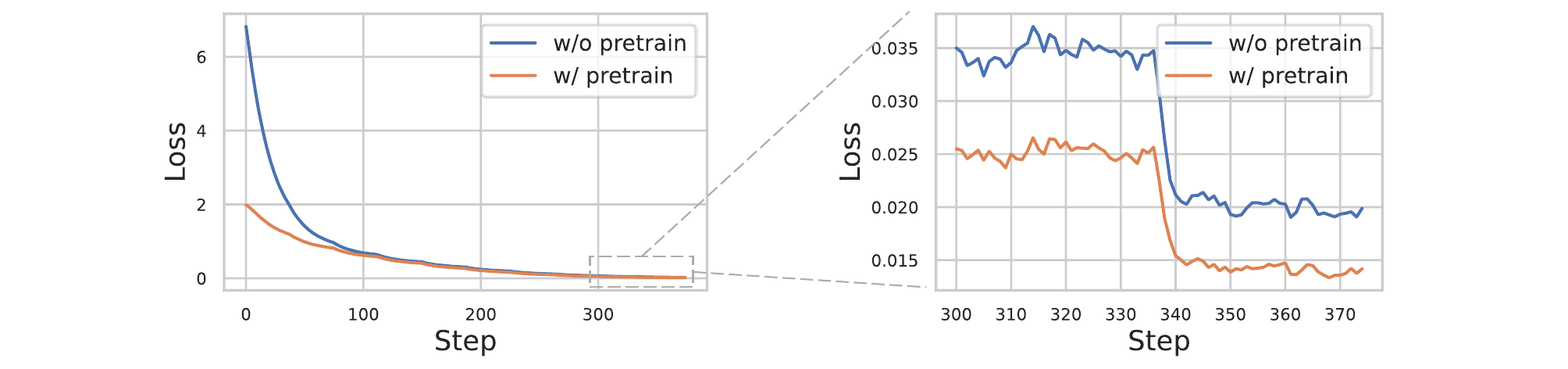}
  \caption{\textbf{Comparison of the loss function curves with and without pretraining.} Two HME models are fine-tuned on the molecular captioning task; however, one underwent the first stage of pretraining, while the other did not.}
  \label{fig:loss_pretrain}
\end{figure}

\begin{table}[t]
\centering
\caption{\textbf{Performance for HME of different sizes.} HME-Small: The auto-regressive decoder is initialized by Llama-3.2-1B. HME-Medium: The auto-regressive decoder is initialized by Llama-3.2-3B. HME: The auto-regressive decoder initialized by Llama-3.0-8B.}
\label{tab:different-HME}
\footnotesize
\begin{tabular}{llcccccc}
\toprule
\textbf{Task} &\textbf{Model}&\textbf{BLEU-2$\uparrow$}&\textbf{BLEU-4$\uparrow$}&\textbf{ROUGE-1$\uparrow$}&\textbf{ROUGE-2$\uparrow$}&\textbf{ROUGE-L$\uparrow$}&\textbf{METEOR$\uparrow$}\\
\midrule
\multirow{3}{*}{Captioning}
 & HME-Small & 32.23&23.94&39.98&24.25&34.31&35.69 \\
  & HME-Medium & 32.83&24.38&41.36&25.54&35.60&36.99 \\
 & HME & 36.26&27.79&43.01&27.58&37.27&39.02 \\
\midrule
\multirow{3}{*}{General QA}
 & HME-Small & 44.43&35.89&50.76&34.80&46.44&53.69 \\
 & HME-Medium & 51.28&42.42&55.72&39.82&51.29&57.75 \\
 & HME & 54.94 &46.41 & 58.94 & 43.59 & 54.90 &60.15 \\
\bottomrule
\addlinespace[0.1cm]
\end{tabular}
\end{table}

\begin{table}[t]
\centering
\footnotesize
\caption{\textbf{Ablation studies for HME.} w/o SMILES: SMILES is not input to the model. w/o Geometry: Geometric features are not input to the model. w/o Fragment: Fragment tokens are not input to the model. w/o Pretrain: The first-stage pretraining procedure is not implemented. w/o Greedy: Probabilistic sampling is used instead of greedy sampling when decoding.}
\label{tab:ablations-HME}
\begin{tabular}{lcccccc}
\toprule
 \textbf{Model} &\textbf{BLEU-2$\uparrow$}&\textbf{BLEU-4$\uparrow$}&\textbf{ROUGE-1$\uparrow$}&\textbf{ROUGE-2$\uparrow$}&\textbf{ROUGE-L$\uparrow$}&\textbf{METEOR$\uparrow$}\\
\midrule
 HME-Small & 32.23&23.94&39.98&24.25&34.31&35.69 \\
 HME-Small (w/o SMILES)  & 18.61&10.92&29.60&13.96&24.26&23.81 \\
HME-Small (w/o Geometry)  & 11.12&4.92&20.76&6.90&15.68&15.17 \\
HME-Small (w/o Fragment)  & 27.72&19.14&36.55&20.45&30.87&31.87 \\
 HME-Small (w/o Pretrain)  & 29.59&21.57&38.25&22.78&32.84&34.01 \\
HME-Small (w/o Greedy)  & 27.91&19.58&35.75&20.09&30.22&31.29 \\
\bottomrule
\addlinespace[0.1cm]
\end{tabular}
\end{table}

\textbf{Other Ablation Studies.}
As shown in Supplementary~Table.~\ref{tab:ablations-HME}, the results demonstrate the importance of each component in our HME model. The baseline HME-Small achieves the best performance across all metrics, with a BLEU-4 score of 23.94 and a ROUGE-L score of 34.31. Removing different input modalities significantly impacts the model's performance, with the geometric features being the most crucial component---excluding them leads to a dramatic drop in performance (BLEU-4 decreases from 23.94 to 4.92). Similarly, removing SMILES representation and fragment tokens also results in substantial performance degradation, with BLEU-4 scores dropping to 10.92 and 19.14 respectively.

We also investigate the effect of pretraining. As shown in Supplementary~Fig~.\ref{fig:loss_pretrain}, we observe that during the first 100 steps, the model with pretraining achieves a lower loss on downstream tasks compared to the model without pretraining. Furthermore, at convergence, the model with pretraining achieves a lower loss value (0.014 vs. 0.020), which highlights the effectiveness of pretraining in molecule-text alignment. Moreover, as shown in Supplementary~Table~.\ref{tab:ablations-HME}, the first-stage pretraining procedure contributes to the model's performance, as evidenced by a decrease in BLEU-4 score from 23.94 to 21.57 when this stage is removed. These results demonstrate the critical role of pretraining in enhancing the model's ability to model chemical-linguistic space, thereby improving both the learning efficiency and the overall performance.

During next-token prediction, our HME employs a greedy sampling strategy. As shown in Supplementary~Table.~\ref{tab:ablations-HME}, we observe that when probabilistic sampling (e.g., nucleus sampling) was used instead of greedy sampling, the model's performance on text generation tasks decreased. For instance, the BLEU-4 score dropped from 23.94 to 19.58, the ROUGE-1 score decreased from 39.98 to 35.75, and the ROUGE-L score declined from 34.31 to 30.22. This indicates that greedy sampling is more effective in generating molecular description texts, enabling the model to produce more accurate textual outputs.

\subsection*{Supplementary Section~7: Visualization of Molecular Fragments}
As shown in Supplementary~Fig~.\ref{fig:visualized_vocab_graphfp}, we randomly selected 50 fragments from our fragment vocabulary for visualization. This vocabulary was constructed using a subset of 456K molecules from the ChEMBL database and contains 800 entries. It was used in all tasks except for the multi-objective molecular reverse design task. In the case of the multi-objective molecular reverse design task, due to the significantly higher number of atoms and rings in natural product molecules compared to drug-like molecules, we constructed a separate vocabulary of size 800 based on a mixed subset of 100K natural product molecules and 100K drug-like molecules, as shown in Supplementary~Fig.~\ref{fig:visualized_vocab_800}.

\subsection*{Supplementary Section~8: Discussion about Molecular Fragment}
Fragment-Based Drug Design (FBDD) is a pivotal strategy in drug discovery, where the core concept is to first identify favorable molecular fragments and then combine them into lead molecules~\cite{yang2024digfrag,orgmoldesign}. A critical step in FBDD is the fragmentation of molecules. During the development of HME, we explored two types of fragmentation methods: rule-based fragmentation~\cite{degen2008brics,lewell1998recap} and graph-based fragmentation~\cite{kong2022psvae,micam}.

Rule-based fragmentation offers certain advantages, such as ease of interpretation and implementation, but the novelty of the resulting fragments is limited~\cite{yang2024digfrag}. Moreover, as observed in our exploration, these methods produce an excessively large and variable-sized fragment vocabulary, which poses challenges for the application of LMs. Specifically, it becomes difficult to map a fragment to a single token within LMs. Although previous researchers have developed fragment-based molecular linear representations using these methods~\cite{wu2024tsmiles}, these fragments are still not based on vocabulary IDs and the representations often rely on complex grammatical rules. We are concerned that this could lead to significant conflicts with the internal knowledge of LLMs and increase the learning difficulty. Therefore, we did not use these methods in our HME.

Graph-based fragmentation derives a fixed-size fragment vocabulary by mining high-frequency subgraphs (i.e., motifs), and the novelty of these fragments aids in the exploration of unknown chemical spaces~\cite{micam}. By expanding the vocabulary of LLMs, the model gains the ability to understand and generate fragment tokens while avoiding confusion with natural language tokens. Although this approach also has the limitation of Out of Vocabulary~(OOV), it can be easily addressed by adding single-atom fragments to the vocabulary. Therefore, our HME adopts this approach.


However, graph-based molecular fragmentation methods exhibit significant limitations: they focus solely on the topological structure of molecules (i.e., the connectivity between atoms) while neglecting many crucial chemical properties, such as atom types, chirality, and bond types.

Recently, the development in the field of computer vision has attracted our attention~\cite{ramesh2021dalle}. In the domain of image generation, researchers have developed a framework based on Vector Quantized Variational AutoEncoder~(VQ-VAE)~\cite{van2017vqvae}, which has demonstrated exceptional performance and flexibility. The working principle of this framework involves three stages: first, the encoder of VQ-VAE transforms an image into a finite set of discrete codes (analogous to IDs in a vocabulary); second, an auto-regressive model learns how to generate code sequences; finally, the decoder of VQ-VAE reconstructs the original image from the discrete codes.
Inspired by this technology, we sought to apply VQ-VAE to the field of molecular fragmentation, aiming to simultaneously account for both the topological structure and the chemical properties of molecules. In our experiments, we successfully transformed attributed molecular graphs into discrete codes. However, we encountered significant challenges in the attempt to reconstruct the original molecules from these discrete codes. Additionally, we found that a molecule with $n$ atoms is typically encoded by VQ-VAE into more than $n$ codes, which contradicts our initial goal of using VQ-VAE for fragmentation. In conclusion, whether VQ-VAE can be utilized effectively for molecular fragmentation remains an open question that requires collaborative efforts from the community.



\begin{figure}[t]
  \centering
  \includegraphics[width=0.8\textwidth]{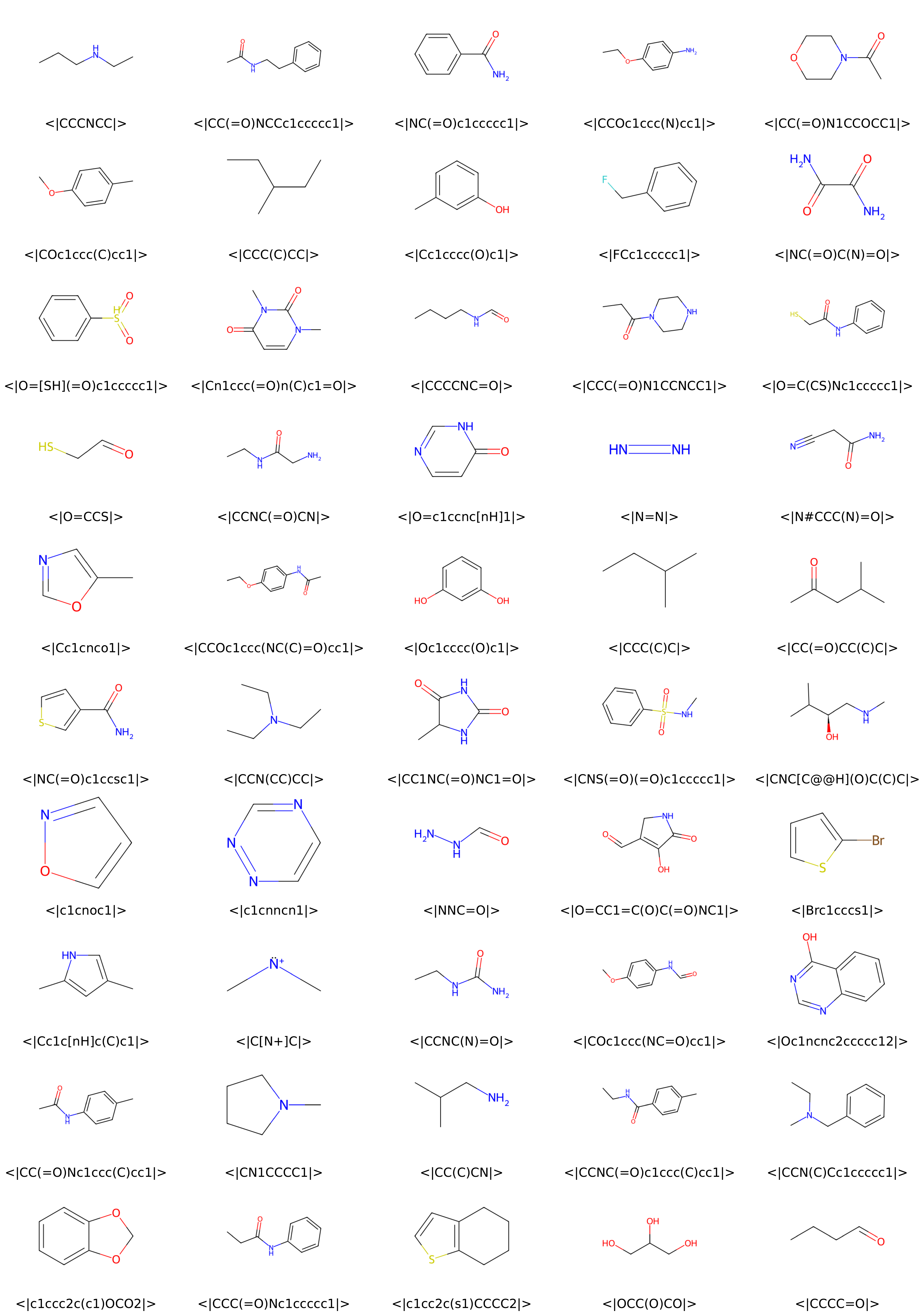}
  \caption{\textbf{Visualization of random 50 fragments in our fragment vocabulary.} The SMILES strings are also provided, with the prefix ``<|” and the suffix ``|>” used to indicate they are fragment tokens.}
  \label{fig:visualized_vocab_graphfp}
\end{figure}

\begin{figure}[t]
  \centering
  \includegraphics[width=0.8\textwidth]{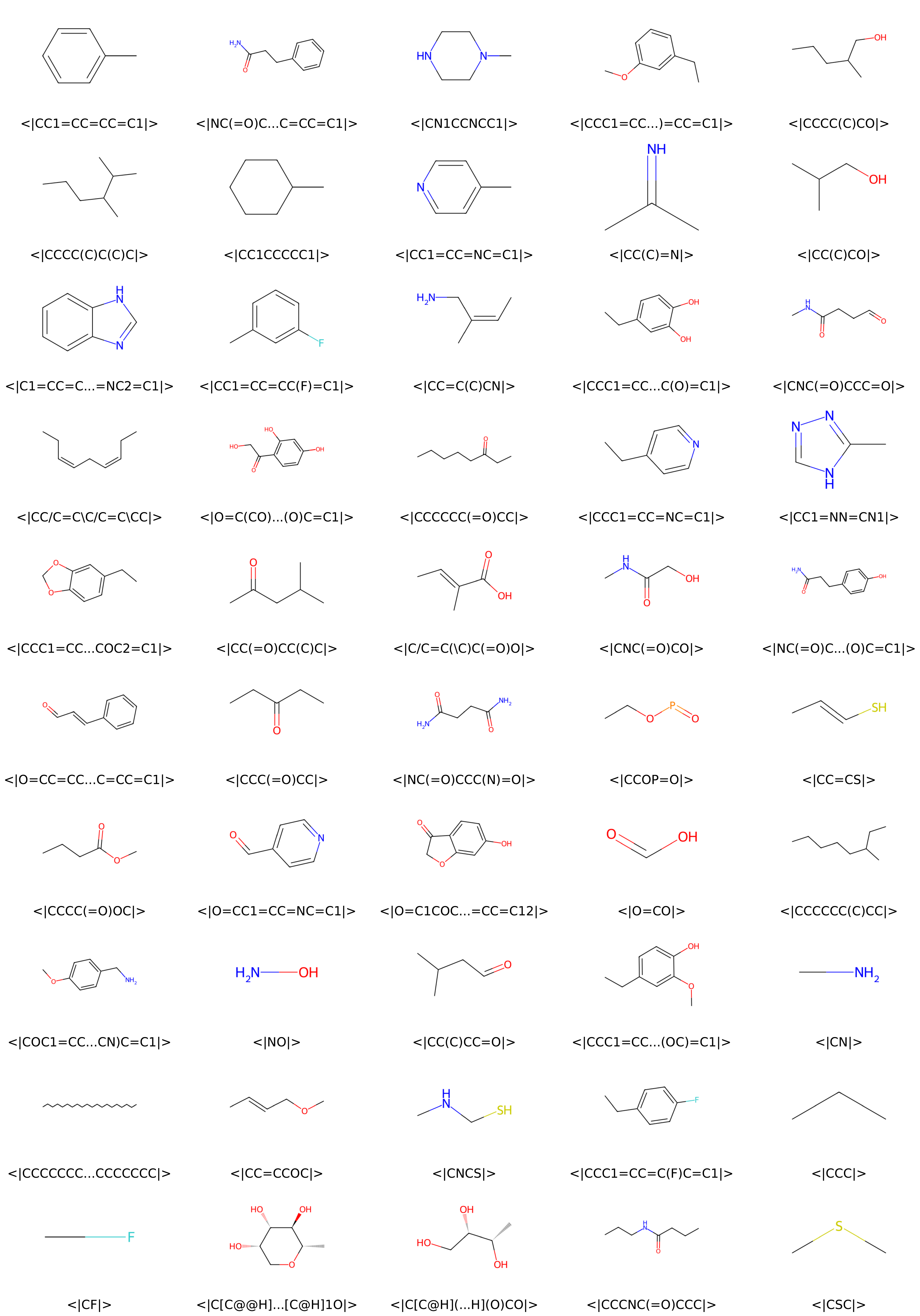}
  \caption{\textbf{Visualization of random 50 fragments in our fragment vocabulary for the multi-objective molecular reverse design task.} The SMILES strings are also provided, with the prefix ``<|” and the suffix ``|>” used to indicate they are fragment tokens.}
  \label{fig:visualized_vocab_800}
\end{figure}

\clearpage

\end{document}